\newif\ifcheckpagelimits
\checkpagelimitstrue
\checkpagelimitsfalse

\ifcheckpagelimits
 
 \documentclass[nofootinbib,prl,aps,twocolumn,showpacs,showkeys,%
 amsmath,amssymb,superscriptaddress,final,reprint,floatfix,longbibliography]{revtex4-1}
 
 \newcommand{\todo}[1]{}
\else
 \documentclass[prl,aps,twocolumn,showpacs,showkeys,%
 amsmath,amssymb,superscriptaddress,final,reprint,floatfix,longbibliography]{revtex4-1}
 \newcommand{\todo}[1]{{\pdfmargincomment[icon=Note,color=pink]{#1}}}
\fi

\usepackage{lineno}
  \usepackage{mathptmx}
\usepackage{subfigure}
\usepackage{dcolumn}
\usepackage{amsmath,amssymb}
\usepackage{bm}
\usepackage{color}
\usepackage{overpic}
\usepackage{latexsym}
\usepackage{epstopdf}
\usepackage[english]{babel}
\usepackage{latexsym}
\usepackage{stmaryrd}

\definecolor{mygrey}{gray}{0.35}
\definecolor{myblue}{rgb}{0.2,0.2,0.8}
\definecolor{myzard}{cmyk}{0,0,0.05,0}
\definecolor{mywhite}{rgb}{1,1,1}
\definecolor{myred}{rgb}{1,0.,0.3}

\usepackage[colorlinks=true,citecolor=myblue,linkcolor=myred,urlcolor=myblue]{hyperref}

 \def\ee{\mathord{\rm e}}

\renewcommand{\ee}{{\rm e}}
\def\beq{\begin{equation}}
\def\eeq{\end{equation}}

\usepackage{graphicx}
\usepackage{soul}
\usepackage{epstopdf}
\usepackage{dsfont}

\newcommand{\ket}[1]{\left\vert #1 \right\rangle}
\newcommand{\bra}[1]{\left\langle #1 \right\vert}
\newcommand{\ketbra}[2]{\ket{ #1}\bra{ #2} }
\newcommand{\bla}[1]{\left( #1 \right)}
\newcommand{\blb}[1]{\left[ #1 \right]}

\def \ket#1{|#1\rangle}
\def \bra#1{\langle#1|}

\def \be{\begin{equation}}
\def \ee{\end{equation}}
\def \ba{\begin{array}}
\def \ea{\end{array}}
\def \bea{\begin{eqnarray}}
\def \eea{\end{eqnarray}}

\renewcommand{\phi}{\varphi}

\usepackage{tabularx,ragged2e,booktabs}
\newcolumntype{C}[1]{>{\Centering}m{#1}}

\begin{document}

\title{Proposal for high-fidelity quantum simulation using hybrid dressed state}
\author{Jianming Cai}
\affiliation{School of Physics, Huazhong University of Science and Technology, Wuhan 430074, China}
\affiliation{Institut f\"{u}r Theoretische Physik, Albert-Einstein Allee 11, Universit\"{a}t Ulm, 89069 Ulm, Germany}
\affiliation{Center for Integrated Quantum Science and Technology, Universit\"{a}t Ulm, 89069 Ulm,
Germany}
\author{Itsik Cohen}
\affiliation{Racah Institute of Physics, The Hebrew University of Jerusalem, Jerusalem 91904, Givat Ram, Israel}
\author{Alex Retzker}
\affiliation{Racah Institute of Physics, The Hebrew University of Jerusalem, Jerusalem 91904, Givat Ram, Israel}
\author{Martin B. Plenio}
\affiliation{Institut f\"{u}r Theoretische Physik, Albert-Einstein Allee 11, Universit\"{a}t Ulm, 89069 Ulm, Germany}
\affiliation{Center for Integrated Quantum Science and Technology, Universit\"{a}t Ulm, 89069 Ulm,
Germany}


\begin{abstract}
A fundamental goal of quantum technologies concerns the exploitation of quantum coherent dynamics for the realisation of novel quantum applications such as quantum computing, quantum simulation, and quantum metrology. A key challenge on the way towards these goals remains the protection of quantum coherent dynamics from environmental noise. 
Here, we propose a concept of hybrid dressed state from a pair of continuously driven systems. It allows sufficiently strong driving fields to suppress the effect of environmental noise, while at the same time being insusceptible to both the amplitude and phase noise in the continuous driving fields. This combination of robust features significantly enhances coherence times under realistic conditions, and at the same time provides new flexibility in Hamiltonian engineering that otherwise is not achievable. We demonstrate theoretically applications of our scheme for noise resistant analog quantum simulation in the well studied physical systems of nitrogen-vacancy centers in diamond and of trapped ions. The scheme may also be exploited for quantum computation and quantum metrology. 
\end{abstract}
\date{\today}
\maketitle

{\it Introduction.---}  Quantum technologies hold the promise for the realisation of a wide variety of applications including quantum computing \cite{Ladd10}, quantum simulation \cite{Nori14,Jaksch14,Cirac12,Grei02,Frie08,Leib02,Kim10}, quantum metrology \cite{Huelga97} and precision measurements \cite{Gio11} as well as quantum sensing \cite{Kohler13,Wra13} to 
name just a few. This motivates the considerable effort that is being invested in the creation of the technological basis for these devices with the ultimate goal of constructing quantum devices that can outperform their classical counterparts. One of the main obstacles on this path is the effect of noise and decoherence due to interaction with an uncontrolled environment,
the effect of which becomes increasingly severe as the number of system components grows. This poses a considerable 
challenge for achieving the quantum control of such systems while maintaining the quantum coherence of the system. Hence noise control is central for the future development of scalable quantum technologies. 

Various theoretical concepts and proposals, e.g. quantum error correction \cite{Shor95,Steane96}, decoherence free subspace \cite{Palma96,Duan97,Zanardi97,Lidar98,Plenio97,Wu2002,Lidar_review}, and dynamical decoupling \cite{Hahn50,Viola98,Pas04,Fan07,Lidar05,Uhrig07,Rabl09,Gordon07,CaiNJP12,Tim11}, have been developed for the suppression or avoidance of quantum decoherence. Each of these methods are best suited for specific scenarios. For example, the decoherence free subspace approach is mainly useful in those cases in which noise exhibits a symmetry \cite{Lidar_review}, which however is not always the case. 
One important example among others that suffer from non-collective noise is the nitrogen-vacancy (NV) center in diamond \cite{Doh13} as a promising physical system for quantum information processing and quantum sensing, the noise of which is dominated by the local nuclear and paramagnetic environment. These noise sources exhibit relatively long memory times and thus can be suppressed using continuous driving. This approach however suffers drawbacks for strong driving where intensity and phase fluctuations may become significant \cite{CaiNJP12}. It is an important observation, that for decoupling fields that are derived from the same source, these fluctuations will be strongly correlated. Hence, we propose to exploit this feature by pairing two driven systems into hybrid dressed states. The energy gap arising from continuous driving protects quantum coherence in non-Markovian environments and makes it insusceptible to inhomogeneous local environment noise, while the pairing approach makes it robust against the driving fluctuation. In comparison with dynamical decoupling in decoherence free subspace, see e.g. \cite{Lidar_review,Wu2002}, the present idea of hybrid dressed state provides new flexibility in the engineering of effective Hamiltonian. With detailed numerical studies, we demonstrate that the present scheme can prolong coherence times by more than two orders of magnitude under realistic conditions. We exploit hybrid dressed states to implement quantum gates and quantum simulation with significantly improved fidelity, e.g. using NV centers in diamond, which thus provides a route towards a large-scale reliable quantum simulation. It is also feasible to verify the idea with other physical systems such as trapped ions as will be demonstrated here by detailed simulations.

{\it The proposal.---} We start from single two-level systems with the basic Hamiltonian
\begin{equation}
H_z^k=\frac{\omega_0}{2} \sigma_z^k, \label{eq:basic}
\end{equation}
where $\sigma_{\alpha}$ ($\alpha=x,y,z$) represent Pauli operators. The coherence of a two-level system, e.g. NV center in diamond, would inevitably be affected by local environmental noise, such as magnetic field and/or charge fluctuation, which can be described by $\Delta_{\alpha}^k=\blb{\delta_{\alpha}^k(t)/2}\sigma_{\alpha}^k$. Generally, $\delta_{\alpha}^k(t)$ are independent for different two-level systems. The dressed states induced by a continuous driving field
\begin{equation}
H_x^k=\frac{\Omega}{2}\bla{ e^{-i\omega_0 t }\sigma_+^k+e^{i\omega_0 t }\sigma_-^k},
\end{equation}
are protected from noise by an energy penalty \cite{Rabl09,Gordon07}, where $\sigma_{\pm}^k=\frac{1}{2}(\sigma_{x}^k\pm \sigma_{y}^k)$, $\Omega$ and $\omega_0$ are the amplitude and frequency of the resonant driving field respectively. In the interaction picture with respect to $H_z^k$, the effective system Hamiltonian becomes
\begin{equation}
H_{e}^k=\frac{\Omega}{2}\sigma_x^k,
\label{eq:He}
\end{equation}
whose eigenstates $\{\ket{{\uparrow_x,\downarrow_x}}\}$ can be used to represent a (simple) dressed spin-$\frac{1}{2}$ that is protected by the continuous driving as long as the power spectrum of the noise $\delta_{z}^k(t)$ at the frequency $\Omega$ is negligible. The phase fluctuations in the continuous driving field have two effects, one to flip the dressed states, which would be suppressed by the energy penalty as well, while the other leads to effective amplitude fluctuations \cite{SI}. The main decoherence source is therefore due to the amplitude fluctuations of continuous driving fields and the residual effect of environment noise due to a finite continuous driving field strength, which causes dephasing of dressed states. It is important to note that if, as is often the case, the decoupling fields are derived from the same source, their fluctuations will be correlated - a feature that we will exploit now. 

\begin{figure}[t]
\begin{center}
\hspace{-0.2cm}
\includegraphics[width=8cm]{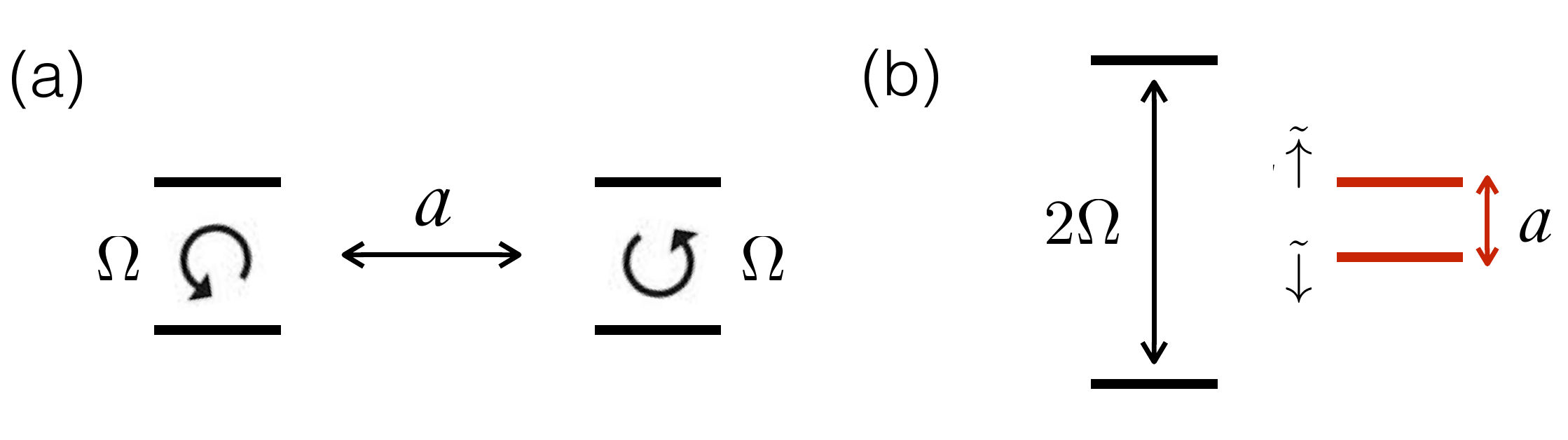}
\end{center}
\caption{(Color online) Hybrid dressed states from a pair of continuously driven quantum systems. {\bf (a)} The effective spin-$\frac{1}{2}$ system is carried by a pair of two-level systems (with Ising-like coupling strength $a$) under continuous driving (with a Rabi frequency $\Omega$). {\bf (b)} The corresponding energy structure of hybrid dressed states. 
}
\label{fig:model}
\end{figure}

To overcome these challenges and making use of the correlated nature of the driving fluctuations, we propose to use a pair of dressed states to carry one effective spin-$\frac{1}{2}$, see Fig.\ref{fig:model}, represented by two of the eigenstates of $H_e^{k}=H_{e}^{k_a}+H_{e}^{k_b}$ as 
\begin{eqnarray}
\ket{\tilde{\uparrow}} &\equiv& \frac{1}{2}\bla{\ket{{\uparrow_x}{\downarrow_x}}_{ab}+\ket{{\downarrow_x}{\uparrow_x}}_{ab}},\\
\ket{\tilde{\downarrow}} &\equiv& \frac{1}{2}\bla{\ket{{\uparrow_x}{\downarrow_x}}_{ab}-\ket{{\downarrow_x}{\uparrow_x}}_{ab}}.
\end{eqnarray}
The states $\ket{\tilde{\uparrow}}$ and $\ket{\tilde{\downarrow}}$ that represent the hybrid dressed spin-$\frac{1}{2}$ are separated from the other two eigenstates $\ket{{\uparrow_x }{\uparrow_x} }$ and $\ket{{\downarrow_x }{\downarrow_x }}$ by an energy gap $ \Omega$. The dephasing noise would induce the spin flip as $\sigma_z \ket{{\uparrow}_x} \rightarrow \ket{{\downarrow}_x}$ and vice versa, the effect of which on the hybrid dressed spin-$\frac{1}{2}$ can be suppressed by tuning the amplitude of the driving field $\Omega$ to sufficiently high values so that the relevant noise power spectra become insignificant. The amplitude fluctuations in the driving field, which are described as $\Delta_{\Omega}^{k}=\blb{\delta_{\Omega}(t)/2}\sigma_x^{k}$, would however cause dephasing of the simple dressed spin-$\frac{1}{2}$  as $\Delta_{\Omega}^{k} \ket{{\uparrow_x}} =\blb{\delta_{\Omega}(t)/2}\ket{{\uparrow_x}}$ and $\Delta_{\Omega}^{k} \ket{{\downarrow_x}} =-\blb{\delta_{\Omega}(t)/2}\ket{{\downarrow_x}}$. Thanks to the correlations in the amplitude fluctuations, the choice of hybrid dressed spin-$\frac{1}{2}$ states, renders the effective spin-$\frac{1}{2}$ instead insusceptible to such fluctuations, because $\blb{\delta_{\Omega}(t)(\sigma_x^{(k,l)_a}+ \sigma_x^{(k,l)_b})}\ket{\tilde{\uparrow}}(\ket{\tilde{\downarrow}})=0$. This allows us to apply strong driving while being immune to their fluctuations.

\begin{figure}[t]
\begin{center}
\hspace{-0.2cm}
\includegraphics[width=8.8cm]{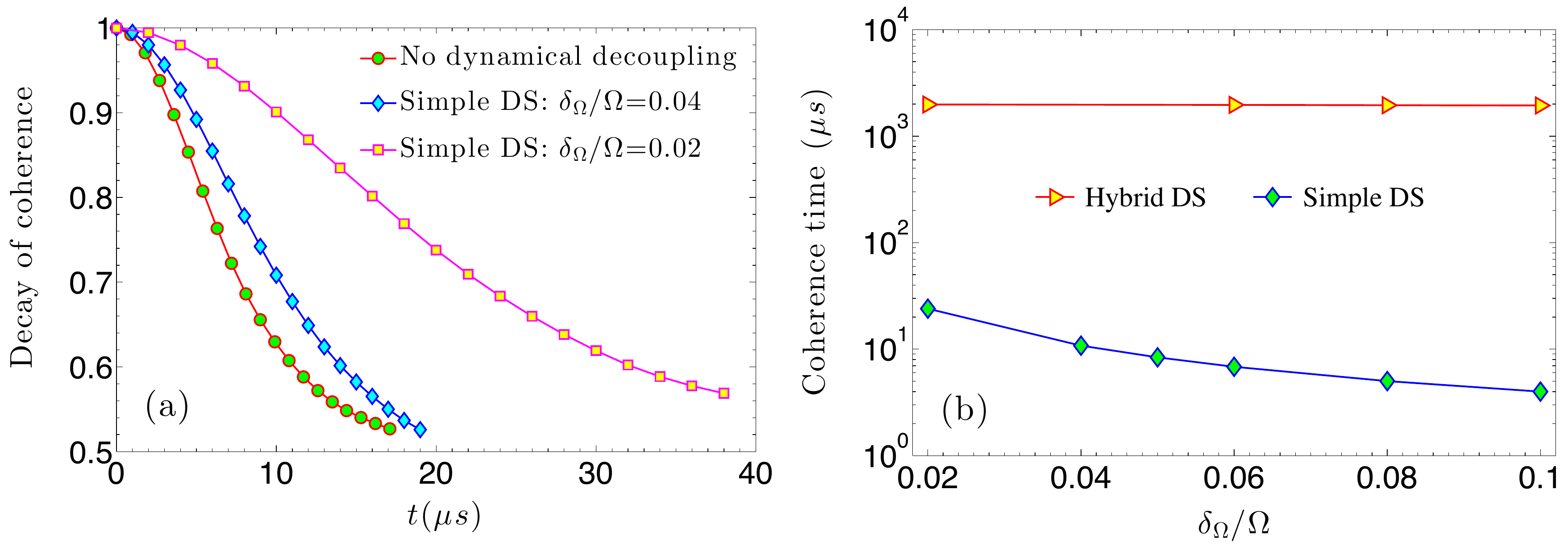}
\end{center}
\caption{(Color online) Stability of hybrid dressed states. (a) The decay of coherence as a function of time $t$ for the cases of no dynamical decoupling and for simple dressed spin (simple DS) with driving fluctuation of $\delta_{\Omega}/\Omega=0.02, 0.04$. (b) The coherence time of hybrid dressed spin (hybrid DS) as a function of the driving fluctuation as compared with a simple dressed spin. The other parameters are: the noise amplitude $\delta=0.2\mbox{MHz}$, and the correlation time is $\tau_c=20\mu s$, the driving amplitude $\Omega=3.5\mbox{MHz}$, and the phase fluctuates in $[-2^{\circ}, 2^{\circ} ]$.}
\label{fig:lifetime}
\end{figure}

In Fig.\ref{fig:lifetime}, we plot the decay of quantum coherence of hybrid dressed states under environment noise and driving fluctuations. The noise (fluctuation) $\delta(t)$ is modeled as an Ornstein-Uhlenbeck process \cite{Daniel96} with the correlation function $C(t) =\langle \delta(0)\delta(t)\rangle= \delta \exp(-|t|/\tau_c)$, where $\delta$ is the noise amplitude and $\tau_c$ is the noise correlation time. We initially prepare a coherent superposition of two basis states, namely $\ket{\psi(t=0)}=\sqrt{\frac{1}{2}}\bla{\ket{\tilde{\uparrow}}+\ket{\tilde{\downarrow}}}$ for hybrid dressed spin, and $\ket{\psi(t=0)}=\sqrt{\frac{1}{2}}\bla{\ket{{\uparrow}}+\ket{{\downarrow}}}$ for a simple dressed spin. We characterize quantum coherence by $f(t)=\bra{\psi(0)}\rho(t)\ket{\psi(0)}$, and define coherence time as $T$ when $f(T)=\frac{1}{2}\bla{1+e^{-1}}$. It can be seen from Fig.\ref{fig:lifetime}(a) that the fluctuation of the continuous driving field indeed would severely limit its efficiency. We demonstrate in Fig.\ref{fig:lifetime}(b) that hybrid dressed states are very robust against the driving fluctuation and can prolong the coherence time by more than two orders of magnitude even for relatively large amplitude and phase fluctuations. 

We illustrate the application of this basic principle by considering two spins, such as NV centers in nanodiamonds as well as shallowly implanted NV-centers for sensing applications. We stress that the same principle is applicable for other physical systems where noise is local and strong driving fields are necessary for decoupling the systems from high level noise. With the ability to address individual spins \cite{Dolde13}, we first prepare the initial dressed states $\ket{{\uparrow}_x}\ket{{\downarrow}_x}$ with microwave-$\frac{\pi}{2}$ pulses, where $\ket{{\uparrow}_x}=\sqrt{\frac{1}{2}}\bla{\ket{{\uparrow}}+\ket{{\downarrow}}}$ and $\ket{{\downarrow}_x}=\sqrt{\frac{1}{2}}\bla{\ket{{\uparrow}}-\ket{{\downarrow}}}$. We drive both spins with the same Rabi frequency and consider the interaction between spins given by $
H_{zz}=a \mathbf{s}_z^{(1)}\otimes \mathbf{s}_z^{(2)}$, where $\mathbf{s}_z^{(1,2)}$ represent the spin-$\frac{1}{2}$ operator in the subspace spanned by $\{\ket{0},\ket{-1}\}$  \cite{Dolde13}. The interaction for time $\tau$ evolves the system into $\ket{\psi(\tau)}=\cos(\frac{a\tau}{4})\ket{{\uparrow}_x}\ket{{\downarrow}_x}-i\sin(\frac{a\tau}{4})\ket{{\downarrow}_x}\ket{{\uparrow}_x}$. The robustness of hybrid dressed states allows us to measure inter-spin interactions and thus the distance between (adjacent) spins with a high precision. This is similar in spirit to the experiment with trapped ions \cite{Ozeri14nature}. There, however, correlated noise dominated while local noise suffered by individual system is always non-collective and needs to be preliminarily addressed by dynamical decoupling. We are thus able to prepare a maximally entangled hybrid dressed state with a very high fidelity, even when the required interaction time significantly exceeds the spin coherence times \cite{SI}. Such a robust hybrid dressed state can also be exploited to construct a sensitive gradient magnetometer based on a pair of dressed spins, e.g. a pair of linked nanodiamonds \cite{Albrecht14}.

\begin{figure}[t]
\begin{center}
\hspace{-0.05cm}
\includegraphics[width=8.5cm]{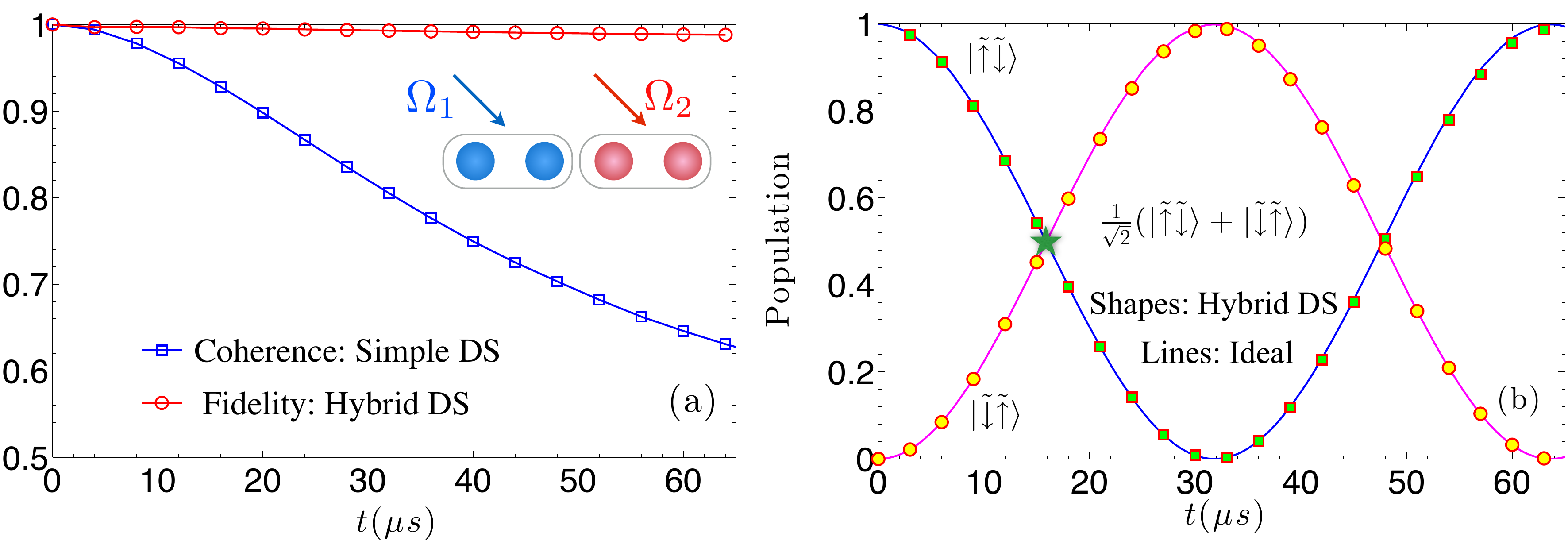}
\end{center}
\caption{(Color online) The robustness of coherent coupling between two hybrid dressed spins (hybrid DS). {\bf (a)} The state fidelity based on hybrid DS with resonant driving (namely $\theta_1=\theta_2=\frac{\pi}{2}$) under realistic noise and driving fluctuation, with respect to the ideal entangling evolution, remains high for the time while the coherence of individual simple dressed spin (simple DS) decays rapidly. {\bf (b)} The coherent oscillation of state population during the entangling evolution. The maximally entangled hybrid dressed state is generated at time $t=\pi/2g_{k,k+1}$ (green star). The relevant parameters are: the noise amplitude $\delta=0.04$\mbox{MHz}, the driving amplitudes $\Omega_{1}=1\mbox{MHz}$, $\Omega_2=3\mbox{MHz}$ with a relative fluctuation $\delta_{\Omega_1}/\Omega_1=\delta_{\Omega_2}/\Omega_2=0.02$; and the nearest-neighbor coupling is $a=(2\pi)20\mbox{kHz}$ (corresponding to a distance of $\sim 15$ nm). The noise and fluctuation correlation time is $\tau_c=20\mu s$.}
\label{fig:ising-coupling}
\end{figure}

{\it Tunable coherent coupling and quantum simulation.---} For the purpose of quantum simulation, it is necessary to engineer {\it tunable} coherent interaction between hybrid dressed states. We consider two pairs of spins under continuous driving, that interact with each other via $H_{k,l}=a_{k,l}\mathbf{s}_z^{k}\otimes\mathbf{s}_z^{l}$ \cite{Dolde13}. With appropriately tuning the amplitude $\Omega$ and frequency ($\omega=\omega_0-\Delta$) of the continuous dressing field, one can obtain the following general form of dressed states $\ket{{\uparrow_{\theta}}}=\cos(\frac{\theta}{2})\ket{{\uparrow}}+\sin(\frac{\theta}{2})\ket{{\downarrow}}$ and $\ket{{\downarrow_{\theta}}}=\sin(\frac{\theta}{2})\ket{{\uparrow}}-\cos(\frac{\theta}{2})\ket{{\downarrow}}$, where $\tan{(\theta)}=\Omega/\Delta$. To make the hybrid dressed states robust against noise and driving fluctuation, we introduce continuous decoupling fields with Rabi frequencies $\Omega'_{k}$ acting on the simple dressed states $\ket{{\uparrow_{\theta}/\downarrow_{\theta}}}$, so that the protection part Hamiltonian is $H_P=\sum_{k}\frac{\Omega'_{k}}{2}\sum_{\alpha}\sigma_{x}^{k_{\alpha}}$. Assuming $\vert \Omega'_k\vert,\vert \Omega'_k \vert, \vert \Omega'_k-\Omega'_l\vert \gg a_{k_{\alpha},l_{\beta}}$ ($\alpha,\beta=a,b$), we obtain from the original spin-spin interaction the following effective interaction Hamiltonian between two hybrid dressed spin-$\frac{1}{2}$, which are carried by ($k_{a},k_{b}$) and ($l_{a},l_{b}$) respectively \cite{SI}, 
\begin{equation}
H=H_P+ \sum_{\mbox{\tiny$\begin{array}{c}
k>l;\alpha,\beta=a,b
\end{array}$}} \bla{J^{x}_{k_{\alpha},l_{\beta}}\sigma_{x}^{k_{\alpha}}\sigma_{x}^{l_{\beta}}+J^{z}_{k_{\alpha},l_{\beta}}\sigma_{z}^{k_{\alpha}}\sigma_{z}^{l_{\beta}}}\label{eq:hdd_coupling}
\end{equation}
where the Pauli operators are written in the basis $\ket{{\uparrow_{\theta}}}, \ket{{\downarrow_{\theta}}}$, and the coupling strength $J^{x}_{k_{\alpha},l_{\beta}}=a_{k_{\alpha},l_{\beta}} \sin(\theta_k)\sin(\theta_l)/2$, $J^{z}_{k_{\alpha},l_{\beta}}=a_{k_{\alpha},l_{\beta}} \cos(\theta_k)\cos(\theta_l)$. This leads to the flexibility in tuning effective spin-spin interactions by controlling the driving parameters. The system Hamiltonian written in the basis of hybrid dressed states is \cite{SI}
\begin{equation}
H=H_P+\bla{h_{k}Z_k+h_{l}Z_l}-g_{k,l}X_{k}X_{l}; \label{eq:effective_coupling}
\end{equation}
where the effective spin-$\frac{1}{2}$ operators are defined as $X\ket{\tilde{\uparrow}}=\ket{\tilde{\downarrow}}, X\ket{\tilde{\downarrow}}=\ket{\tilde{\uparrow}},
Z\ket{\tilde{\uparrow}}=\ket{\tilde{\uparrow}}, Z\ket{\tilde{\downarrow}}=-\ket{\tilde{\downarrow}}$. The local energy is given by $h_{k}=J^{z}_{k_{a},k_{b}}$, and the coupling strength is $g_{k,l}=(J^{x}_{k_a,l_b}+J^{x}_{l_a,k_b})-(J^{x}_{k_a,l_a}+J^{x}_{k_b,l_b})$. In the particular case $\Delta=0$, namely $\theta=\frac{\pi}{2}$, the dressing field also acts the role of dynamical decoupling, thus we have $H_{k,l}= \bla{a_{k,l}/2}{\bla{\sigma_{x}^{k}\sigma_{x}^{l}+\sigma_{y}^{k}\sigma_{y}^{l}}}$ \cite{SI}, which allows to efficiently realize high-fidelity entangling gates between hybrid dressed spins. Assuming equal (intra- and inter- pair) nearest-neighbor interactions $a$ and non-nearest-neighbor couplings as $a_{k,l}=a|k-l|^{-3}$, the effective nearest-neighbor coupling is $g_{k,k+1}\approx 0.78 a$, while $h_{k}=h_{k+1}$. In Fig.\ref{fig:ising-coupling}, it can be seen that the fidelity of maximal entanglement generation at time $t=\pi/2g_{k,k+1}$ is high (above $99\%$) even in the case where the noise/fluctuation is stronger than the coupling strength.

\begin{figure}[t]
\begin{center}
\hspace{-0.3cm}
\includegraphics[width=8.6cm]{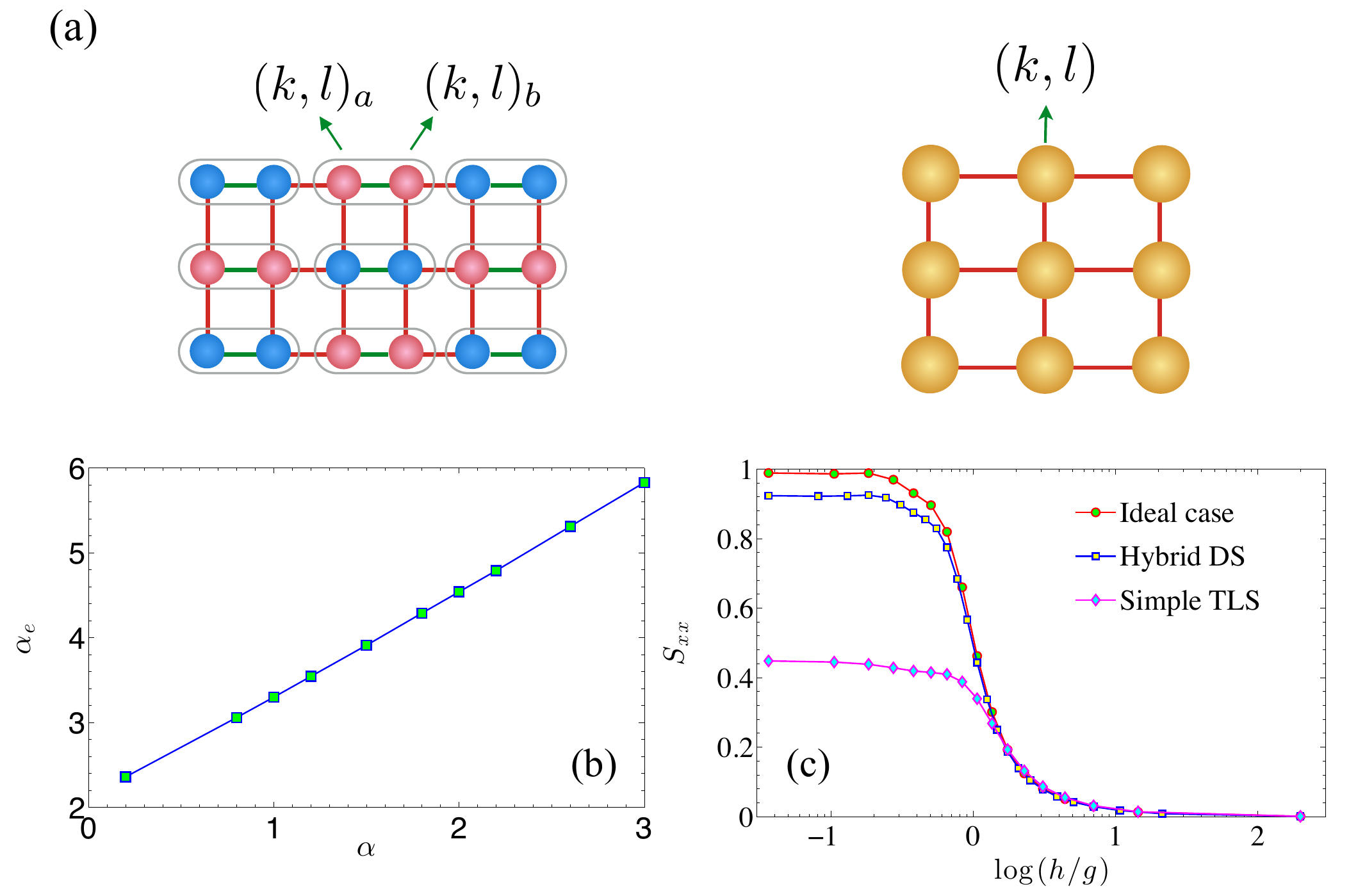}
\end{center}
\caption{(Color online) Quantum simulation using hybrid dressed state. {\bf (a)}  A two-dimensional lattice quantum simulator for the simulation of an interacting spin models. The hybrid dressed spin-$\frac{1}{2}$ of the index $(k,l)$ is represented by driven two-level systems $(k,l)_a$ and $(k,l)_b$. {\bf (b)} The effective spin-spin interaction range $\alpha_{e}$ as a function of the original interaction range $\alpha$. {\bf (c)} The spin structure factor $S_{xx}=\sum_{k,l>k}\exp{(-iq(l-k))} \langle \sigma_x^{(k)} \sigma_x^{(l)} \rangle$ with $q=0$ achieved via adiabatic preparation for a one-dimensional eight-site Ising chain using hybrid dressed spins (hybrid DS) and simple two-level systems (simple TLS). The parameters are $g(t)=a_0\bla{\frac{t}{T}}$, $h(t)=a_0-g(t)$ with $a_0=(2\pi) 40 \mbox{kHz}$ and $T=80\mu s$. 
}
\label{fig:ising-model}
\end{figure}

The generalization of coupling in Eq.(\ref{eq:effective_coupling}) to more two-level systems on a regular lattice, see Fig.\ref{fig:ising-model}(a), e.g. self-assembled nanodiamond lattice \cite{Albrecht14} and trapped ion lattice \cite{Britton12,Berprl11,Bernjp12}, thus leads to hybrid dressed states protected robust simulation of quantum Ising model
$H_{\mbox{s}}=\sum h_{k,l}Z_{(k,l)}-\sum g_{(k,l),(k',l')} X_{(k,l)} X_{(k',l')}$. The coherence protection provided by $H_P$ is independent of the parameters of the simulating Hamiltonian in $H_{\mbox{s}}$. Long-range dipolar interaction, namely the interaction $\sigma_{z}^{k_{\alpha}}\sigma_{z}^{l_{\beta}}$ for $k\neq l$ in Eq.(\ref{eq:hdd_coupling}), should be effectively eliminated, as it would cause the leakage out of the protected subspace. With $K$ periodically spatial alternating driving amplitudes ($[\Omega_1,\Omega_2,\cdots,\Omega_K]^n$) for a one-dimensional chain so that $\vert \Omega_k-\Omega_l\vert \gg g_{k,l}$, see the inset of Fig.\ref{fig:ising-coupling}(a) for the case of $K=2$, the unwanted interactions can be suppressed up to the $(2K-1)$-neighbor coupling. For example, the residual interaction is $\sim 0.008 (0.003) a$ with $K=3 (4)$ for $r^{-\alpha}$ ($\alpha=3$) scaling long range interactions. Analogous techniques can be employed for the general case of two-dimensional lattice. For one-dimensional chain, the effective coupling strength scales as $g_{k,l}\propto {\vert k-l\vert}^{-\alpha_e}$ where $\alpha_e \approx 2.07+1.24\alpha$, see Fig.\ref{fig:ising-model}(b), where $\alpha$ is the original interaction range. This enables us to achieve a shorter interaction length that otherwise is not possible \cite{Islam13}. Furthermore, as we mentioned, the dependence of the interaction on the driving basis, see Eq.(\ref{eq:hdd_coupling}), provide additional flexibility in engineering the simulating Hamiltonian parameters. In contrast to the digital fashion of quantum simulation in decoherence free subspace \cite{Monz09}, the present analog quantum simulation will save precious time in case of simulating a large spin system.

To show the performance of hybrid dressed states in adiabatic quantum simulation, we calculate the spin structure factor  $S_{xx}=\sum_{k,l>k}\exp{(-iq(l-k))} \langle \sigma_x^{(k)} \sigma_x^{(l)} \rangle$ with $q=0$ for a one-dimensional Ising chain. In Fig.\ref{fig:ising-model} (c), it can be seen that the transition of the order parameter $S_{xx}(q=0)$ across the critical point becomes much smoother due to the decoherence effect of noise. One would expect that it will be difficult to observe quantum phase transition as the noise level and/or the system size increases. In contrast, we show in Fig.\ref{fig:ising-model}(c) that, the effect of noise and driving fluctuation is much suppressed in a quantum simulator with the same number of hybrid dressed spins. We remark that the present scheme is also feasible for the study of quantum quench dynamics, which has attracted intensive interest recently, as it is related to the fundamental problem of thermalization in closed many-body quantum systems \cite{Cardy07,Rigol08,Riera12,Trotzky12}. In particular, high-fidelity quantum simulation may help to distinguish local thermalization from environment induced equilibrium \cite{Riera12,Trotzky12}. 

{\it Experiment implementation.---} The present proposal can be implemented in various physical systems, e.g. NV centers in diamond and trapped ions. For NV centers in diamond, the tunable coherent interactions between hybrid dressed states can be engineered from the natural magnetic dipole-dipole interaction using appropriate microwave drivings \cite{SI}. The proof-of-principle verification of the ideas can also be achieved with trapped ions, in which we can induce the interaction term $\sigma_x^a \sigma_x^b$ in the dressed state basis by applying counter-propagating Raman beams for generating the detuned red sideband transition, additionally we drive the carrier dressing transition for shielding the magnetic noise \cite{SI,Porras04,Zhu06,Ber12}. In order to induce the transverse field for the simulated Hamiltonian $\sigma^a_z\sigma^b_z$, we can introduce an additional rotated term which transforms the Ising coupling to the XXZ Hamiltonian \cite{SI,Itsik1,Itsik2}. For that purpose, we use the single addressing ability of the focused co-propagating Raman beams, to drive the dressed states off resonantly \cite{SI}. The feasibility of this approach is carefully verified by our numerical simulation with realistic experiment parameters \cite{SI}.

{\it Conclusion.---} In summary, we propose to implement high-fidelity quantum simulation using hybrid dressed state. Its key advantage is the possibility for applying strong driving to suppress non-collective environment noise while at the same time the inevitable driving fluctuations are suppressed. Additionally, it provides new flexibility in the engineering of tunable coherent coupling. We demonstrate theoretically the applicability of our scheme for noise resistant quantum simulation with NV centers in diamond, but would like to stress that it is sufficiently versatile to apply to a wide variety of physical systems. Our simulations show that a proof-of-principle verification is readily achievable with trapped ions under state-of-the-art experiment capability. The method can be naturally combined with a wide range of applications such as quantum metrology/sensing and quantum computing. Our work thus contributes a new path towards the development of quantum technologies, in which the effect of decoherence is currently a severe obstacle.

{\it Acknowledgements.---} J.-M.C is supported by the startup grant of Huazhong University of Science and Technology and the National Natural Science Foundation of China (grant no.11574103). M. B. P. is supported by the Alexander von Humboldt Foundation, the DFG (SPP1601, and SFB TR21), the EU Integrating Projects SIQS and the EU STREPs EQUAM and DIADEMS, and the ERC Synergy grant BioQ. A. R. is supported by the Israel Science Foundation (grant no.039-8823), the European commission (STReP EQUAM), EU Project DIADEMS, the Marie Curie Career Integration Grant (CIG) IonQuanSense(321798) and the Niedersachsen-Israeli Research Cooperation Program.

\newpage 

\onecolumngrid

\section{Supplementary Information}

\renewcommand{\thefigure}{S\arabic{figure}}
\setcounter{figure}{0}
\renewcommand{\theequation}{S.\arabic{equation}}
\setcounter{equation}{0}

\subsection*{Noise analysis of hybrid dressed state } 

We consider two-level systems as described by the following Hamiltonian
\begin{equation}
H_z^k=\frac{\omega_0}{2} \sigma_z^k,
\end{equation}
where $\sigma_{\alpha}$ ($\alpha=x,y,z$) represent Pauli operators, noise can occur along all three axes $\Delta_{\alpha}^k=\blb{\delta_{\alpha}^k(t) /2}\sigma_{\alpha}^k$. Given the usual case of $\omega_0 \gg \delta_{x,y}^k$, the effect of transverse noise is suppressed by the energy gap $\omega_0$. Thus the main decoherence comes from the longitudinal noise
\begin{equation}
\Delta_{z}^k=\blb{\delta_{z}^k(t)/2} \sigma_{z}^k,
\end{equation}
in the form of dephasing. 
\begin{figure}[b]
\begin{center}
\hspace{-0.35cm}
\includegraphics[width=6.5cm]{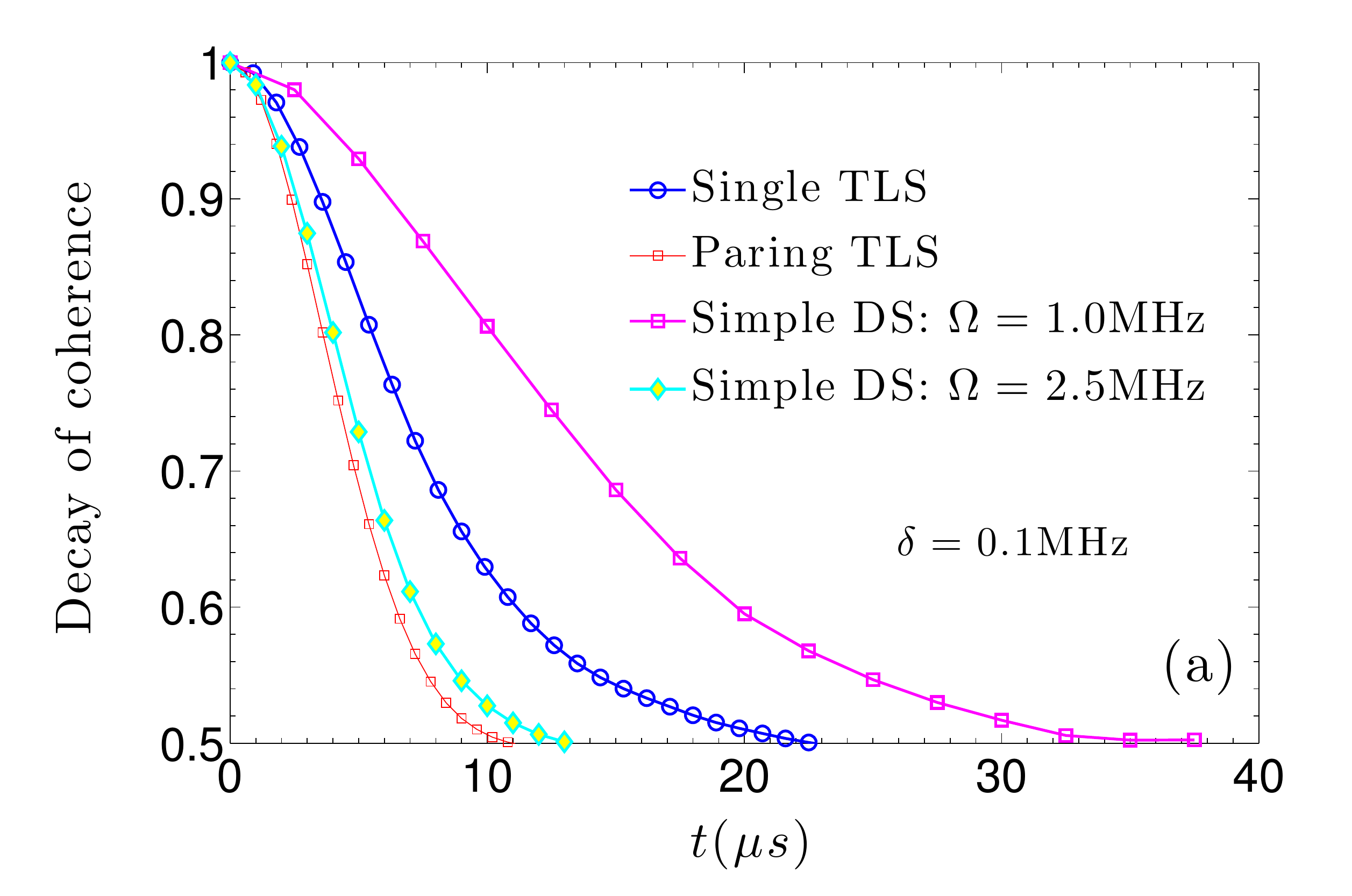}
\hspace{0.2cm}
\includegraphics[width=6.5cm]{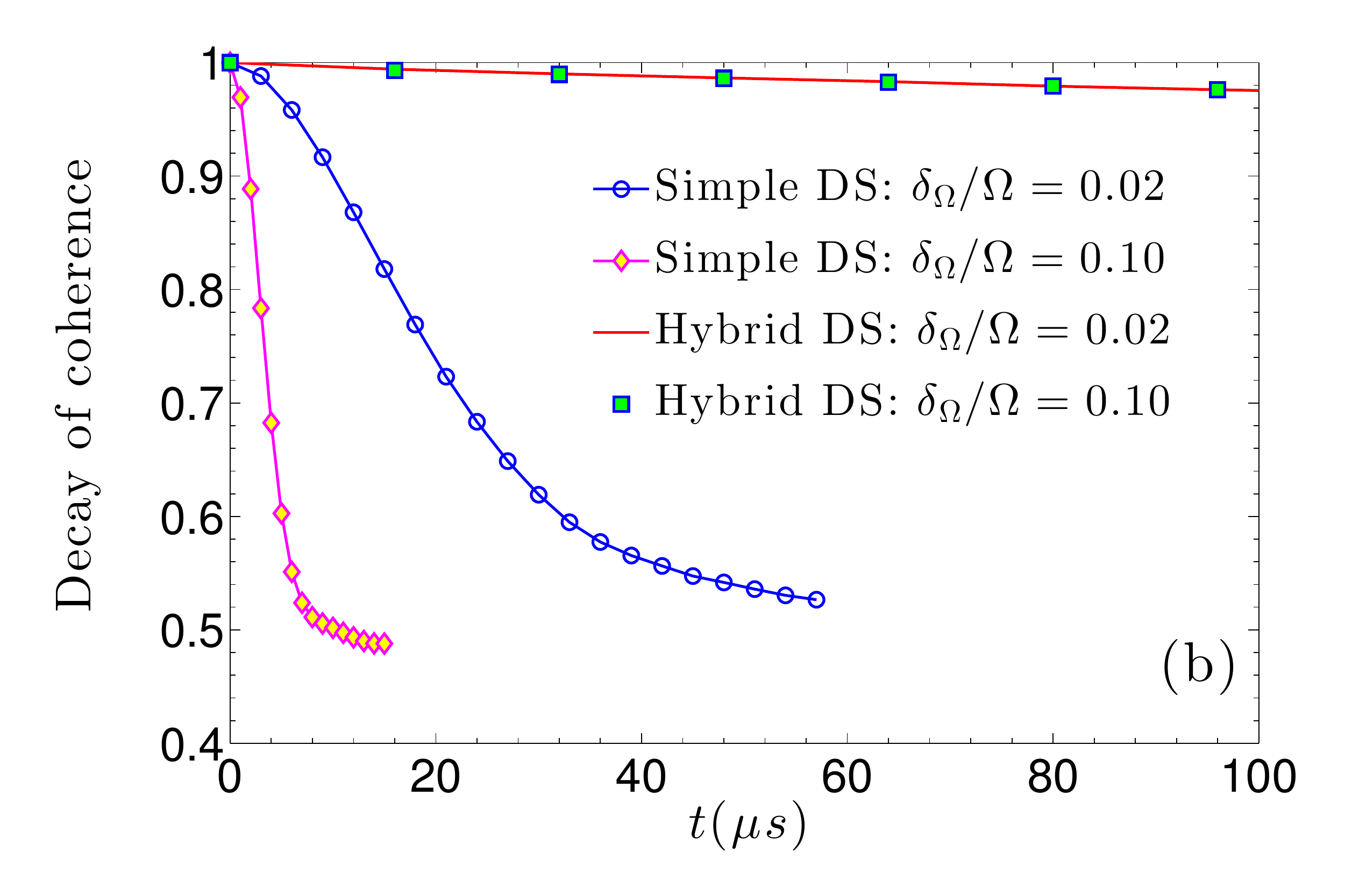}
\end{center}
\caption{(Color online) Enhanced coherence time of hybrid dressed state. (a) The decay of coherence as a function of time for a single two-level system (TLS), and a pair of two-level systems, as compared with simple dressed spin (simple DS) with driving amplitude $\Omega$ and relative fluctuation $\delta_{\Omega}/\Omega=0.1$. (b) The decay of coherence as a function of time for hybrid dressed spin (hybrid DS). The driving amplitude is $\Omega=3.5\mbox{MHz}$. 
}
\label{fig:lifetime-SI}
\end{figure}
In the case of local environment noise, it is not helpful to use a pair of two-level systems to encode an effective spin-$\frac{1}{2}$ as there is no decoherence free subspace, see Fig.~\ref{fig:lifetime-SI}(a). One can introduce dressed state by applying a continuous driving field with amplitude and phase fluctuation
\begin{equation}
\tilde{H}_x^k=\blb{\frac{\Omega}{2} +\frac{\delta _{\Omega}(t)}{2}}\blb{ e^{-i\omega_0 t - i\delta_{p}(t)}\sigma_+^k+e^{i\omega_0 t + i\delta_{p}(t)}\sigma_-^k},
\end{equation}
where $\delta_{\Omega}(t)$ and $\delta_{p}(t)$ represent the amplitude and phase fluctuation of the driving field. The effect of $\Delta_{z}^k$ is suppressed by the driving field, as long as its power spectra is negligible at the frequency of $\Omega$. In the interaction picture with respect to $H_z^k$, the effective Hamiltonian including driving field fluctuations is
\begin{eqnarray}
\tilde{H}_e^k&=&\blb{\frac{\Omega}{2} +\frac{\delta _{\Omega}(t)}{2}}\blb{ \cos\delta_{p}(t) \sigma_x^k+\sin \delta_{p}(t) \sigma_y^k}\\
&=&\frac{\Omega}{2} \sigma_x^k+\frac{\tilde{\delta}_{\Omega}(t)}{2} \sigma_x^k+\blb{\frac{\Omega}{2} +\frac{\delta _{\Omega}(t)}{2}}\sin \delta_{p}(t) \sigma_y^k, \label{eq:si-phase-1}
\end{eqnarray}
where
\begin{equation}
\tilde{\delta}_{\Omega}(t)=\delta_{\Omega}(t) \cos\delta_{p}(t)+\Omega\blb{ \cos\delta_{p}(t)-1}. \label{eq:si-phase}
\end{equation}
The phase fluctuation $\delta_{p}(t)$ in Eq.(\ref{eq:si-phase-1}) shows two effects. First, it leads to additional fluctuation in the amplitude of driving, and the effective total amplitude fluctuation is given by Eq.(\ref{eq:si-phase}). Secondly, it also translates into the flip error ($\sigma_y^k$) of dressed states, which is suppressed if the fluctuation is small such that $\sin \delta_{p}(t) \ll 1$. Thus, decoherence mainly comes from the amplitude fluctuation of driving described as follows
\begin{equation}
\tilde{H}_e^k=\frac{\Omega}{2} \sigma_x^k+\frac{\tilde{\delta}_{\Omega}(t)}{2} \sigma_x^k. \label{eq:si-phase-effective}
\end{equation}
Generally, the amplitude fluctuation becomes larger as the driving power increases. By increasing the driving field amplitude, on the one hand, it is beneficial for the suppression of environment noise, on the other hand, the accompanying increase in the driving field fluctuations would diminish the overall performance, as shown Fig.~\ref{fig:lifetime-SI}(a-b). The essence of the hybrid dressed state is using a pair of two-level systems under identical driving to carry one single effective spin-$\frac{1}{2}$ as
\begin{eqnarray}
\ket{\tilde{\uparrow}} &\equiv& \frac{1}{2}\bla{\ket{{\uparrow_x}{\downarrow_x}}_{ab}+\ket{{\downarrow_x}{\uparrow_x}}_{ab}} \label{eq:dressup}\\
\ket{\tilde{\downarrow}} &\equiv& \frac{1}{2}\bla{\ket{{\uparrow_x}{\downarrow_x}}_{ab}-\ket{{\downarrow_x}{\uparrow_x}}_{ab}},\label{eq:dressdown}
\end{eqnarray}
Because the following conditions are satisfied
\begin{eqnarray}
\blb{\tilde{\delta}_{\Omega}(t)(\sigma_x^{(k,l)_a}+ \sigma_x^{(k,l)_b})}\ket{\tilde{\uparrow}}&=&0,\\
\blb{\tilde{\delta}_{\Omega}(t)(\sigma_x^{(k,l)_a}+ \sigma_x^{(k,l)_b})}\ket{\tilde{\downarrow}}&=&0,
\end{eqnarray}
hybrid dressed states are thus robust not only to environment noise but also to the amplitude and phase fluctuation in driving. 

In our numerical simulation, we model the environment noise $\delta(t)$ and the amplitude fluctuations of the driving fields $\delta_{\Omega}(t)$ as an Ornstein-Uhlenbeck process \cite{Daniel96_SI} with the correlation function $C(t) =\langle \delta(0)\delta(t)\rangle= \delta \exp(-|t|/\tau_c)$, where $\delta$ is the noise amplitude and $\tau_c$ is the noise correlation time. We also include the fluctuating phase $\delta_{p}(t)$ in our model where it takes a random value uniformly distributed in $\blb{-\delta_{p}^0, \delta_{p}^0}$. It can be seen from Fig.~\ref{fig:lifetime-SI}(b) that coherence times of hybrid dressed state are significantly prolonged, and is not susceptible to the fluctuations the continuous driving field. The scaling of coherence times is shown in Fig.2 of main text.\\

\begin{figure}[h]
\begin{center}
\hspace{-0.05cm}
\includegraphics[width=11.6cm]{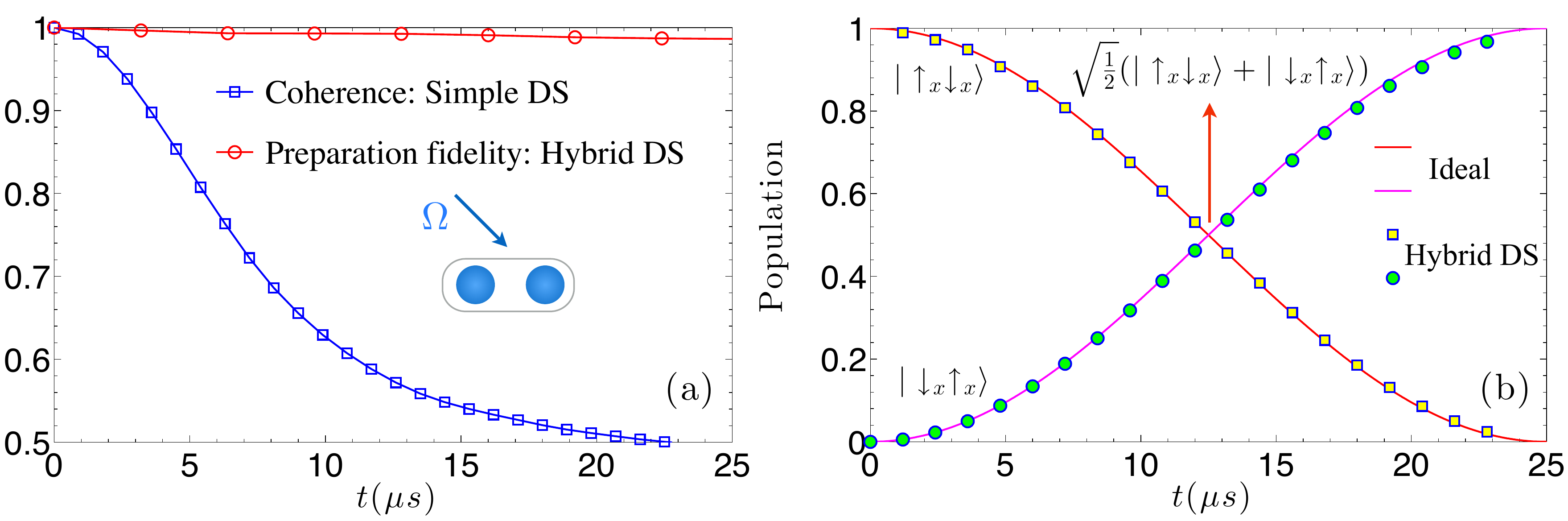}
\end{center}
\caption{(Color online) Robust preparation of hybrid dressed states. {\bf (a)} The state fidelity during the system evolution under realistic noise and driving fluctuation, with respect to the ideal entangling evolution as described in Eq.(\ref{eq:spin-spin_interaction-SI}) remains high for the time while the decay of coherence for a simple dressed spin (simple DS) happens in a much shorter time scale. {\bf (b)} The coherent oscillation of hybrid dressed spin (hybrid DS) state population during the evolution. The relevant parameters are as follows: the noise amplitude is $\delta=0.1$\mbox{MHz}, the amplitude of driving field is $\Omega=3.5\mbox{MHz}$ with a relative fluctuation is $\delta_{\Omega}/\Omega=0.02$; and the nearest-neighbor coupling is $g=(2\pi)20\mbox{kHz}$. The noise and fluctuation correlation time is $\tau_c=20\mu s$.}
\label{fig:ising-coupling-si}
\end{figure}

\subsection*{Robust preparation of hybrid dressed states} 

We consider a pair of NV centers with addressable electronic transition frequencies \cite{Dolde13_SI} to prepare (entangled) hybrid dressed states. We drive both NV centers with the same Rabi frequency, the interaction between NV centers is then given by 
\begin{equation}
H_{S}=\frac{\Omega}{2}\blb{\mathbf{s}_x^{(1)}+\mathbf{s}_x^{(2)}}+a \mathbf{s}_z^{(1)}\otimes \mathbf{s}_z^{(2)}, \label{eq:spin-spin_interaction-SI}
\end{equation} 
where $\mathbf{s}_x^{(1,2)}, \mathbf{s}_z^{(1,2)}$ represent the spin-$\frac{1}{2}$ operators in the subspace spanned by $\{\ket{0},\ket{-1}\}$.  Initially we prepare $\ket{\psi(0)}=\ket{{\uparrow}_x}\ket{{\downarrow}_x}$ with microwave-$\frac{\pi}{2}$ pulses acting on the ground spin sublevels of NV centers, where $\ket{{\uparrow}_x}=\sqrt{\frac{1}{2}}\bla{\ket{0}+\ket{-1}}$ and $\ket{{\downarrow}_x}=\sqrt{\frac{1}{2}}\bla{\ket{0}-\ket{-1}}$. After an interaction time $\tau$, the system evolves into the state $\ket{\psi(\tau)}=\cos(\frac{a\tau}{4})\ket{{\uparrow}_x}\ket{{\downarrow}_x}-i\sin(\frac{a\tau}{4})\ket{{\downarrow}_x}\ket{{\uparrow}_x}$. The state preparation fidelity is severely limited by local magnetic noise, particularly when the spin-spin coupling is even weaker than noise, namely $a\leq \delta$. In Fig.\ref{fig:ising-coupling-si}, we simulate the state preparation of (entangled) hybrid states, and shows a high fidelity ($>99\%$) under realistic noise and driving fluctuation.

\subsection*{Derivation of tunable interaction between hybrid dressed states with NV centers}  
The  tunability of Hamiltonian parameters is necessary for quantum simulation. In order to achieve tunable interaction between NV centers from the spin-spin interaction  $H_{kl}=a_{k,l}{\mbox{s}_{z}^{k}\otimes \mbox{s}_{z}^{l}}$, we use a microwave field with Rabi frequency $\Omega_m$, and frequency detuning $\Delta_m$, the effective Hamiltonian in the interaction picture is
\begin{eqnarray}
H_x^m=\frac{\tilde{\Omega}_m}{2} \blb{\sin\bla{\theta_m}\sigma_x^m+\cos\bla{\theta_m} \sigma_z^m}, \label{eq-si:hxm}
\end{eqnarray}
where $\tilde{\Omega}_m=\sqrt{\Omega_m^2+\Delta_m^2}$ and $\tan{(\theta_m)}=\Omega_m/\Delta_m$. The above Hamiltonian can be diagonalized as
\begin{equation}
H_x^m=\frac{\tilde{\Omega}}{2}\bla{  \ketbra{{\uparrow}{(\theta_m)}} {{\uparrow}{(\theta_m)}} -\ketbra{{\downarrow}{(\theta_m)}} {{\downarrow}{(\theta_m)}}},\label{eq-si:Hxm_d}
\end{equation}
where the corresponding eigenstates are
\begin{eqnarray}
\ket{{\uparrow}{(\theta_m)}}&=&\cos{(\frac{\theta_m}{2})}\ket{{\uparrow}}+\sin{(\frac{\theta_m}{2})}\ket{{\downarrow}},\\
\ket{{\downarrow}{(\theta_m)}}&=&\sin{(\frac{\theta_m}{2})}\ket{{\uparrow}}-\cos{(\frac{\theta_m}{2})}\ket{{\downarrow}}.
\end{eqnarray}
The spin-spin interaction $H_{k,l}$ can then be written in the basis of $\{\ket{{\uparrow}(\theta_m)},\ket{{\downarrow}(\theta_m)}\}$ ($m=k,l$) as 
\begin{eqnarray}
H_{k,l}=a_{k,l}\blb{\cos{(\theta_k)}\sigma_{z}^{k}+\sin{(\theta_k)}\sigma_{x}^{k}}\bigotimes \blb{\cos{(\theta_l)}\sigma_{z}^{l}+\sin{(\theta_l)}\sigma_{x}^{l}}.\label{eq-s17}
\end{eqnarray}
Under the assumption that $a_{k,l}\ll \tilde{\Omega}_k, \tilde{\Omega}_l$, the cross interaction terms $\sigma_{x}^{k} \sigma_{z}^{l}$ and $\sigma_{z}^{k} \sigma_{x}^{l}$ can be neglected, and thus in the interaction picture with respect to Eq.(\ref{eq-si:Hxm_d}) we obtain
\begin{eqnarray}
H_{k,l}=a_{k,l}\blb{\cos{(\theta_k)} \cos{(\theta_l)} \sigma_{z}^{k}\sigma_{z}^{l}+\frac{1}{2}\sin{(\theta_k)}\sin{(\theta_l)}\bla{\sigma_{x}^{k}\sigma_{x}^{l}+\sigma_{y}^{k}\sigma_{y}^{l}}}. \label{eq-si:sameR}
\end{eqnarray}
We note that this microwave field is exploited to achieve tunable interaction between NV centers, thus we do not require that it should be identical on all NV centers. For example, one can use $\tilde{\Omega}_{m_a}$ and $-\tilde{\Omega}_{m_b}$ for two NV centers that carry a single hybrid spin-$\frac{1}{2}$, and in this way the interaction becomes 
\begin{eqnarray}
H_{k,l}=a_{k,l}\blb{\cos{(\theta_k)} \cos{(\theta_l)} \sigma_{z}^{k}\sigma_{z}^{l}+\frac{1}{2}\sin{(\theta_k)}\sin{(\theta_l)}\bla{\sigma_{x}^{k}\sigma_{x}^{l}-\sigma_{y}^{k}\sigma_{y}^{l}}}.\label{eq-si:altR}
\end{eqnarray}
According to the Trotter decomposition, the effective interaction between two NV centers from choosing alternating Rabi frequencies (i.e. Eq.\ref{eq-si:sameR}-\ref{eq-si:altR}) becomes
\begin{eqnarray}
H_{k,l}=a_{k,l}\blb{\cos{(\theta_k)} \cos{(\theta_l)} \sigma_{z}^{k}\sigma_{z}^{l}+\frac{1}{2}\sin{(\theta_k)}\sin{(\theta_l)}\left( {\sigma_{x}^{k}\sigma_{x}^{l}}+f_{k,l}\sigma_{y}^{k}\sigma_{y}^{l}\right)},\label{eq-si:hkl-eff}
\end{eqnarray}
where $f_{k,l}=\blb{1+(-1)^{k-l}}/2$ and we will see that the term $\sigma_{y}^{k}\sigma_{y}^{l}$ can be eliminated in the subspace of hybrid dressed states. Therefore, we obtain the following effective Hamiltonian for hybrid dressed state as
\begin{eqnarray}
H_{k,l}=a_{k,l}\blb{\cos{(\theta_k)} \cos{(\theta_l)} \sigma_{z}^{k}\sigma_{z}^{l}+\frac{1}{2}\sin{(\theta_k)}\sin{(\theta_l)}{\sigma_{x}^{k}\sigma_{x}^{l}}}.\label{eq-si:hkl-eff-xz}
\end{eqnarray}

To achieve robust features against both local magnetic noise and fluctuation in driving fields, we can introduce hybrid dressed states starting from Eq.(\ref{eq-si:Hxm_d}), similar as Eq.(1) in the main text. In the same interaction picture of Eq.(\ref{eq-si:hxm}), the noise and the driving field fluctuation is described by
\begin{equation}
H_{noise}^m=\frac{\delta\Omega_m(t)}{2} \sigma_x^m+\frac{\delta B_m(t)}{2} \sigma_z^m,
\end{equation}
Written in the rotated basis of $\{\ket{{\uparrow}(\theta_m)},\ket{{\downarrow}(\theta_m)}\}$, we obtain
\begin{equation}
H_{noise}^m=\blb{-\frac{\delta\Omega_m(t)}{2}\cos\theta_m+\frac{\delta B_m(t)}{2}\sin
\theta_m }\sigma_x^m+\blb{\frac{\delta\Omega_m(t)}{2}\sin\theta_m+\frac{\delta B_m(t)}{2}\cos\theta_m }. \sigma_z^m. \label{eq-si:Hnoise}
\end{equation}
In the interaction picture with respect to Eq.(\ref{eq-si:Hxm_d}), the flip effect $\sigma_x^m$ in $H_{noise}^m$ (see Eq.\ref{eq-si:Hnoise}) is suppressed, and thus
\begin{equation}
H_{noise}^m\approx\blb{-\frac{\delta\Omega_m(t)}{2}\sin\theta_m+\frac{\delta B_m(t)}{2}\cos\theta_m } \sigma_z^m.\label{eq-si:hnoise}
\end{equation}
In order to suppress this noise, we introduce continuous driving fields, which are presented in the lab frame as follows:
\begin{equation}
\Omega'_m \sigma^m_x\cos (\omega^m_0 + \tilde{\Omega}_m)t - \Omega'_m \sigma^m_x\cos (\omega^m_0 - \tilde{\Omega}_m)t.
\label{decoupling}
\end{equation}
Moving to the interaction picture with respect to the bare energy structure yields 
\begin{equation}
\frac{\Omega'_m}{2} \sigma^m_+ e^{-i  \tilde{\Omega}_m t} -\frac{\Omega'_m}{2} \sigma^m_+ e^{i  \tilde{\Omega}_m t}+h.c = \Omega'_m \sigma^m_y \sin  \tilde{\Omega}_m t,
\end{equation}
in the rotating wave approximation (RWA) assuming $\Omega'_m \ll  \omega^m_0\pm  \tilde{\Omega}_m$. We now transform to the rotated basis and move to the second interaction picture with respect to Eq. \ref{eq-si:Hxm_d}. Therefore, in the RWA assuming $\Omega'_m \ll  \tilde{\Omega}_m$, we obtain the continuous driving fields
\begin{equation}
H_d^m=-\frac{\Omega'_m}{2}\blb{ \ketbra{{\uparrow}{(\theta_m)}} {{\downarrow}{(\theta_m)}} +\ketbra{{\downarrow}{(\theta_m)}} {{\uparrow}{(\theta_m)}}}=-\frac{\Omega'_m}{2}\sigma_x,
\label{second}
\end{equation}
which would then suppress the noise in Eq.(\ref{eq-si:hnoise}). The remaining noise comes from the fluctuation of the driving field
\begin{equation}
H_n^m=-\frac{\delta_{\Omega'}(t)}{2}\sigma_x^m.
\end{equation}
The total effective Hamiltonian is
\begin{eqnarray}
H=H_d^m+H_n^m+H_{k,l} 
\end{eqnarray}
The corresponding dressed states (i.e. the eigenstates of $H_d^m$) are
\begin{eqnarray}
\ket{{\uparrow}_x} &=& \frac{1}{2}\bla{\ket{{\uparrow(\theta_m)}}+\ket{{\downarrow(\theta_m)}}} ,\\
\ket{{\downarrow}_x} &=& \frac{1}{2}\bla{\ket{{\uparrow(\theta_m)}}-\ket{{\downarrow(\theta_m)}}} ;
\end{eqnarray}
and the hybrid dressed states are
\begin{eqnarray}
\ket{\tilde{\uparrow}} &\equiv& \frac{1}{2}\bla{\ket{{\uparrow_x}{\downarrow_x}}_{ab}+\ket{{\downarrow_x}{\uparrow_x}}_{ab}} \label{eq-si:dressup}\\
\ket{\tilde{\downarrow}} &\equiv& \frac{1}{2}\bla{\ket{{\uparrow_x}{\downarrow_x}}_{ab}-\ket{{\downarrow_x}{\uparrow_x}}_{ab}},\label{eq-si:dressdown}
\end{eqnarray}
As we discuss in the main text, the above hybrid dressed states are robust against both local magnetic noise and fluctuations of driving fields. In the following, we will derive the tunable Hamiltonian of coupled NV centers by writing the total effective Hamiltonian in the basis of hybrid spin-$\frac{1}{2}$. For the intra-pair interaction (between NV centers that carry one hybrid dressed spin-$\frac{1}{2}$), because $f_{m_a,m_b}=0$ in Eq.\ref{eq-si:hkl-eff}, is
\begin{eqnarray}
H_{m_a,m_b}=a_{m_a,m_b}\blb{\cos{(\theta_{m_a})} \cos{(\theta_{m_b})} \sigma_{z}^{{m_a}}\sigma_{z}^{{m_b}}+\frac{1}{2}\sin{(\theta_{m_a})}\sin{(\theta_{m_b})}{\sigma_{x}^{{m_a}}\sigma_{x}^{{m_b}}}}.
\end{eqnarray}
This gives the following local Hamiltonian of hybrid dressed spin-$\frac{1}{2}$ 
\begin{equation}
H_{L}^m=J_{m_{a},m_{b}}^{z} Z_{m},
\end{equation}
with
\begin{equation}
J_{m_{a},m_{b}}^{z} =a_{m_a,m_b}\blb{\cos{(\theta_{m_a})} \cos{(\theta_{m_b})}},
\end{equation}
where $Z\ket{\tilde{\uparrow}}=\ket{\tilde{\uparrow}}, Z\ket{\tilde{\downarrow}}=-\ket{\tilde{\downarrow}}$. For the inter-pair interaction between NV centers that carry different hybrid dressed spin-$\frac{1}{2}$, $\sigma_{y}^{k_{\alpha}}\sigma_{y}^{l_{\beta}}$ for ($k\neq l$) will flip the dressed states $\ket{\uparrow_x}_k \leftrightarrow \ket{{\downarrow_x}}_k$, and $\ket{{\uparrow_x}}_l \leftrightarrow \ket{{\downarrow_x}}_l$. Therefore, it can be effectively eliminated if $\Omega'_k,\Omega'_l, \vert \Omega'_k-\Omega'_l\vert \gg a_{k_{\alpha},l_{\beta}}\blb{\sin{(\theta_{k_{\alpha}})} \sin{(\theta_{l_{\beta}})}}$. This explains why we can obtain the effective Hamiltonian for hybrid dressed states in Eq.\ref{eq-si:hkl-eff-xz}. Similarly, $\sigma_{z}^{k_{\alpha}}\sigma_{z}^{l_{\beta}}$ can also be eliminated. This gives the effective interaction between hybrid dress spin-$\frac{1}{2}$ as
\begin{equation}
H_{I}^{k,l}=\blb{\bla{J^{x}_{k_a,l_b}+J^{x}_{l_a,k_b}}-\bla{J^{x}_{k_a,l_a}+J^{x}_{k_b,l_b}}}X_{k}X_{l},
\end{equation}
where
$X\ket{\tilde{\uparrow}}=\ket{\tilde{\downarrow}}, X\ket{\tilde{\downarrow}}=\ket{\tilde{\uparrow}}$.
Thus, we obtain the following tunable effective Hamiltonian
\begin{equation}
H_{s}=H_{P}+\sum_{m} H_{L}^m +\sum_{k,l}H_{I}^{k,l}=\sum_{m} H_d^m+\sum_{m} J_{m_{a},m_{b}}^{z} Z_{m} +\sum_{k,l}\blb{\bla{J^{x}_{k_a,l_b}+J^{x}_{l_a,k_b}}-\bla{J^{x}_{k_a,l_a}+J^{x}_{k_b,l_b}}}X_{k}X_{l},
\end{equation}
an example of which is shown in Eq.(7) of the main text for two hybrid dressed spin-$\frac{1}{2}$.

Note that in order to induce the Trotter decomposition, by alternating  $\tilde{\Omega}_{m_b}$, the driving fields should be changed correspondingly.  Namely, we should alternate the detuned driving field's Rabi frequency $\Omega_{m_b}$ and detuning $\Delta_{m_b}$ (Eq. \ref{eq-si:hxm}), and the Rabi frequency of the continuous driving fields $\Omega'_{m_b}$  (Eq. \ref{decoupling}) as well.

\subsection*{Derivation of the Hamiltonian for trapped atomic ion }
\subsubsection*{Derivation of the $\sigma_x\sigma_x$ interactions.}

To achieve $\sigma_x\sigma_x$ interactions, we apply counter-propagating Raman beams in addition to a resonant carrier transition \cite{Ber12_SI}, as presented in the following Hamiltonian:
\begin{eqnarray}
H^{i,n}&=& \nu_n b^\dagger_n b_n\label{phonons} \\
&+&\frac{\omega_0}{2}\sigma_z^i +\omega_e \ket{e}^i\bra{e}\label{bare}\\
&+& \Omega_1\left( \ket{e}^i\bra{1}+h.c\right) \cos \left[ (\omega_e-\frac{\omega_0}{2}-\Delta)t +\eta_{i,n}( b_n^\dagger + b_n) \right] \label{Raman1}\\
&+&\Omega_0\left( \ket{e}^i\bra{0}+h.c\right) \cos \left[ (\omega_e+\frac{\omega_0}{2}-\Delta-\delta)t -\eta_{i,n}( b_n^\dagger + b_n) \right]\label{Raman2}\\
&+&\Omega_c\sigma^i_x\cos \left({\omega_0}t  \right)\label{dressing1}.
\end{eqnarray}
Here, Eq.(\ref{phonons}) is the vibration Hamiltonian that is described by the creation (annihilation) operators $ b_n^\dagger ,(b_n) $  for the $n^{th}$ vibrational mode. Eq.(\ref{bare}) is the bare energy structure where $\omega_e \ket{e}^i\bra{e}$ is the virtually excited state, in the optical regime, through which the Raman transitions (Eq.\ref{Raman1}-\ref{Raman2}) are operated. $\sigma^i_z= \ket{1}^i\bra{1}- \ket{0}^i\bra{0}$ and $\sigma^i_x= \ket{1}^i\bra{0}+ \ket{0}^i\bra{1}$, are the Pauli matrices that operate on the qubit states of the $i^{th}$ ion, and $\eta_{n,i}$ is the Lamb-Dicke parameter of the $i^{th}$ ion and the $n^{th}$ vibrational mode. Finally, Eq.(\ref{dressing1}) is the dressing field's term.

The Hamiltonian in the interaction picture with respect to the bare energy structure (Eq. \ref{bare}) becomes:
\begin{eqnarray}
H_I^{i,n}&=& \nu_n b^\dagger_n b_n +\frac{\Omega_1}{2} \left(\ket{e}^i\bra{1} e^{i \left[\Delta t -\eta_{i,n}( b_n^\dagger + b_n) \right]}+h.c \right)+\frac{\Omega_0}{2} \left( \ket{e}^i\bra{0}e^{i\left[(\Delta+\delta)t +\eta_{i,n}( b_n^\dagger + b_n) \right]}+h.c\right)   \\
&+&\frac{\Omega_c}{2}\sigma^i_x \label{dressed},
\end{eqnarray}
where we used the RWA assuming $\Omega_1,\Omega_0 \ll \omega_e-\omega_0$ and $\Omega_c \ll \omega_0$. In the second order of perturbation theory, we obtain the following effective Hamiltonian:
\begin{equation}
\begin{split}
H^{i,n}_{eff}= \nu_n b^\dagger_n b_n +\frac{\Omega_c}{2}\sigma^i_x -\frac{\Omega_1\Omega_0}{4\Delta} \left[ \sigma_-^i e^{-i\delta t} e^{-i 2\eta_{i,n}( b_n^\dagger + b_n)} +h.c \right],\label{eff_xx}
\end{split}
\end{equation}
where we neglected A.C. Stark shifts. Since the A.C Stark shifts are described by $\sigma_z^i$ operators, they can be eliminated by adjusting the detunings of the Raman beams, and in any case, their first order contribution is suppressed by the dressing field $(\Omega_c/2)\sigma_x^i$ (Eq.\ref{dressed}). 

In the next step, we expand the exponent of the displacement operator to the first order in the Lamb-Dicke parameter.  The zeroth order, gives rise to a detuned carrier transitions, resulting in an A.C Stark shift that is suppressed from the above reasons. 
In the rotating frame of the vibrational modes (Eq.\ref{phonons}) and the dressed state energy gap (Eq.\ref{dressed}), the operator $\sigma_-^i = (1/2)\left(\sigma^i_x- i\sigma_y^i\right)$ is effectively reduced to $\sigma^i_x/2$ since $\sigma_y^i$ is suppressed if $\delta-\nu_n \ll \omega_c$; thus the first order expansion yields the S\o rensen-M\o lmer Hamiltonian:
\begin{equation}
\begin{split}
H^{i,n}_{I_2}=i\frac{\eta_{n,i}\Omega_1\Omega_0}{4\Delta}  \sigma_x^i b_n^\dagger e^{-i(\delta-\nu_n) t}   +h.c,
\end{split}\label{SM Interaction}
\end{equation}
which results in the effective Ising Hamiltonian,
\begin{equation}
\begin{split}
H_{{eff}_2}=\left(\frac{\Omega_0\Omega_1}{4\Delta}\right)^2\sum_{i,j,n} 
\frac{\eta_{n,i}\eta_{n,j}}{\delta-\nu_n}  \sigma_x^i  \sigma_x^j, \end{split} \label{SM Hamiltonian}
\end{equation}
 in the second order of perturbation assuming $\frac{\eta_{n,i}\Omega_1\Omega_0}{4\Delta} \ll \delta-\nu_n$. 
 
\subsubsection{Derivation of the $\sigma^a_z\sigma^b_z$ interactions.}

\begin{figure}[t]
\begin{center}
\includegraphics[width=15cm]{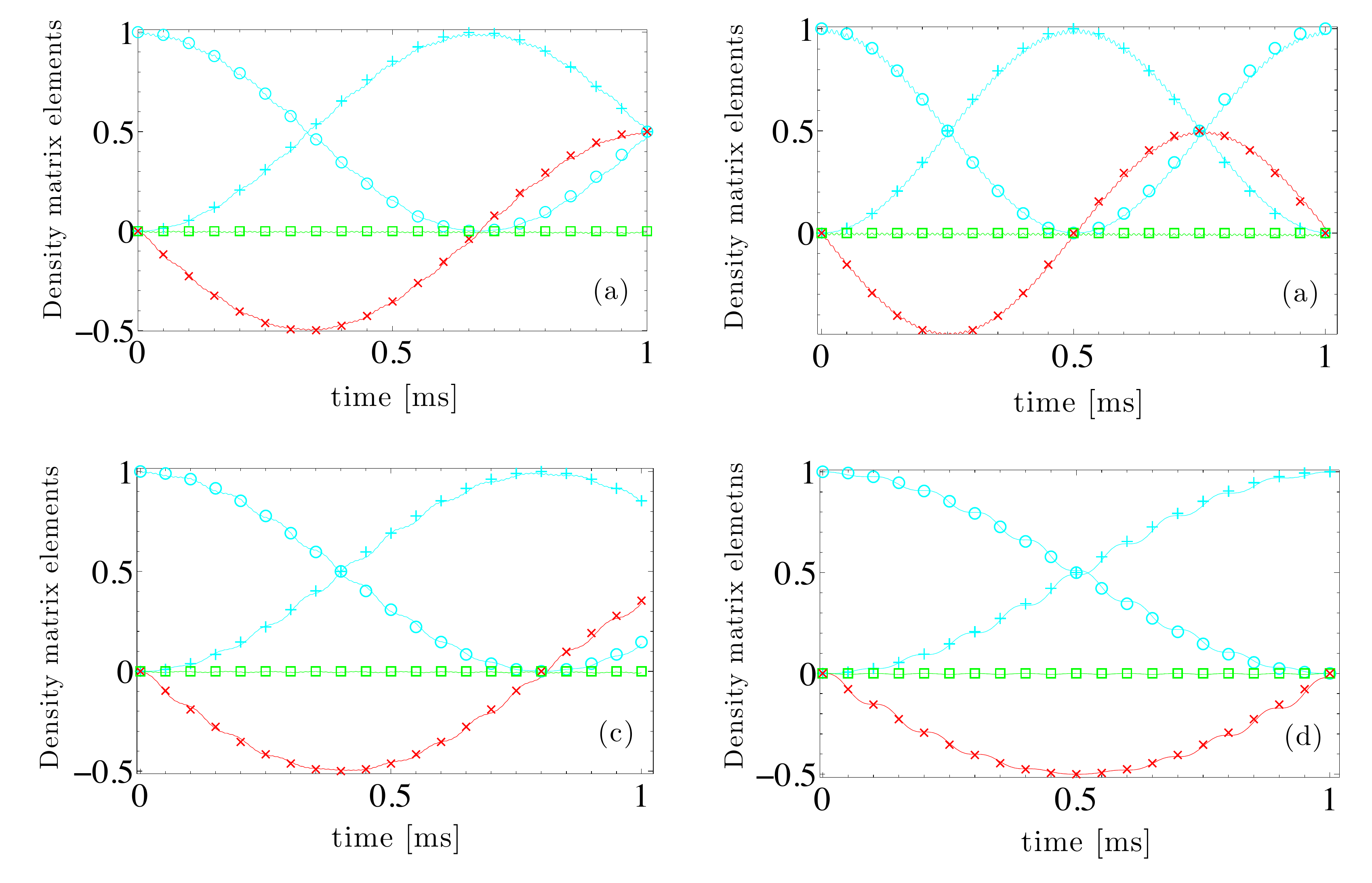}
\end{center}
\caption{(Color online) Simulation of the XXZ Hamiltonian for trapped ions. We simulate the XXZ Hamiltonian for two ions and one vibrational mode using Eq.(\ref{eff_xx}-\ref{eff2}) (solid lines), and show the equivalence to Eq.(\ref{XXZ Hamiltonian}) (markers) for different $\theta$. We initialize the system with $\ket{{\uparrow}{(\theta)}{\downarrow}{(\theta)}}$ and the motional ground state$\ket{n=0}$, and measure the reduced density matrix's elements of: $\rho_{\ket{{\uparrow}{(\theta)}{\downarrow}{(\theta)}}\bra{{\uparrow}{(\theta)}{\downarrow}{(\theta)}}}$ (cyan circles), $\rho_{\ket{{\downarrow}{(\theta)}{\uparrow}{(\theta)}}\bra{{\downarrow}{(\theta)}{\uparrow}{(\theta)}}}$  (cyan pluses),   $\Im \left(\rho_{\ket{{\downarrow}{(\theta)}{\uparrow}{(\theta)}}\bra{{\uparrow}{(\theta)}{\downarrow}{(\theta)}}}\right)$  (red exes), and $\Re \left(\rho_{\ket{{\downarrow}{(\theta)}{\uparrow}{(\theta)}}\bra{{\uparrow}{(\theta)}{\downarrow}{(\theta)}}}\right)$  (green squares), as usually done when measuring entanglement. For that purpose, we use the following parameters: $ \frac{\Omega_0\Omega_1}{4\Delta}=2\pi \cdot 100$kHz, $\eta=0.05$, $\nu=2\pi \cdot 10$MHz, $\delta=2\pi \cdot10.1$MHz, $\Omega_c=2\pi\cdot 2$MHz. 
 {\bf (a)} For $\theta=0$ we use $\Omega^z_m=0$ and $\Delta_m=2\pi\cdot 2$MHz . 
 {\bf (b)} For $\theta=\pi/4$ we use $\Omega^z_m=2\pi\cdot 4.98$kHz and $\Delta_m=2\pi\cdot4.99$kHz. 
 {\bf (c)} For $\theta=\pi/3$ we use $\Omega^z_m=2\pi\cdot4.99 $kHz and $\delta_m=2\pi\cdot2.88$kHz. {\bf (d)} For $\theta=\pi/2$ we use $\Omega^z_m=2\pi\cdot 5$kHz and $\Delta_m= 0$.}
\label{fig-si:ion_sim}
\end{figure}

To achieve the $\sigma^i_z\sigma^j_z$ interactions, we can introduce an additional rotated term, which breaks the Ising coupling to obtain the XXZ Hamiltonian \cite{Itsik1,Itsik2}, similarly to the derivation for the NV centers (Eq.\ref{eq-si:hxm}). 
For realizing the this method we apply co-propagating focused Raman beams, thus having single addressing ability. The driving fields in the lab frame are presented as
\begin{eqnarray}
& & \Omega^m_1\left( \ket{e}^i\bra{1}+h.c\right) \cos \left[ (\omega_e-\frac{\omega_0}{2}-\Delta_2)t +\eta_{i,n}( b_n^\dagger + b_n) \right] \label{Raman'1}\\
&+&\Omega^m_0\left( \ket{e}^i\bra{0}+h.c\right) \cos \left[ (\omega_e+\frac{\omega_0}{2}-\Delta_2-\delta_m)t +\eta_{i,n}( b_n^\dagger + b_n) \right]\label{Raman'2}.
\end{eqnarray}
After moving to the interaction picture with respect to the bare energy structure (Eq.\ref{bare}), and using the RWA where we assume $\Omega^m_1,\Omega^m_0 \ll \omega_e-\omega_0$, we obtain
\begin{eqnarray}
\frac{\Omega^m_1}{2} \left(\ket{e}^i\bra{1} e^{i \left[\Delta_2 t -\eta_{i,n}( b_n^\dagger + b_n) \right]}+h.c \right)+\frac{\Omega^m_0}{2} \left( \ket{e}^i\bra{0}e^{i\left[(\Delta_2+\delta_m)t -\eta_{i,n}( b_n^\dagger + b_n) \right]}+h.c\right).
\end{eqnarray}
In the second order of perturbation theory, this yields
\begin{equation}
\begin{split}
-\frac{\Omega^m_1\Omega^m_0}{4\Delta_2} \left( \sigma_+^i e^{i\delta_m t} +h.c \right),\label{eff2}
\end{split}
\end{equation}
which is detuned from the dressed state energy structure (Eq.\ref{dressed}) by $\Delta_m =\Omega_c - \delta_m$. Now, if we move to the interaction picture with respect to $\left(\delta_m/2\right)\sigma^m_x$, the only term of Eq.\ref{eff2} that preserves energy and thus survives in the RWA, is $(\Omega^z_m/2) \sigma_z $, with $\Omega^z_m=-\frac{\Omega^m_1\Omega^m_0}{4\Delta_2}$. Thus, together with $\left(\Delta_m/2\right)\sigma^m_x$, which is left from the dressed state energy gap (Eq.\ref{dressed}), we obtain the rotated term needed for breaking the Ising coupling into the XXZ one, similarly to Eq.\ref{eq-si:hxm}
\begin{equation}
\begin{split}
H_x^m= \frac{\Omega^z_m}{2} \sigma_z+\frac{\Delta_m}{2} \sigma_x= \frac{\tilde{\Omega}_m}{2} \blb{\sin\bla{\theta_m}\sigma_z^m+\cos\bla{\theta_m} \sigma_x^m},
\end{split}
\end{equation}
where $\tilde{\Omega}_m=\sqrt{{\Omega^z_m } ^2+\Delta_m^2}$ and $\tan{(\theta_m)}=\Omega^z_m/\Delta_m$. This Hamiltonian can be diagonalized as
\begin{equation}
H_x^m=\frac{\tilde{\Omega_m}}{2}\bla{  \ketbra{{\uparrow}{(\theta_m)}} {{\uparrow}{(\theta_m)}} -\ketbra{{\downarrow}{(\theta_m)}} {{\downarrow}{(\theta_m)}}},\label{second interaction}
\end{equation}
where the corresponding eigenstates are
\begin{eqnarray}
\ket{{\uparrow}{(\theta_m)}}&=&\cos{(\frac{\theta_m}{2})}\ket{{\uparrow}}+\sin{(\frac{\theta_m}{2})}\ket{{\downarrow}},\\
\ket{{\downarrow}{(\theta_k)}}&=&\sin{(\frac{\theta_m}{2})}\ket{{\uparrow}}-\cos{(\frac{\theta_m}{2})}\ket{{\downarrow}}.
\end{eqnarray}
Eq. \ref{SM Interaction} can then be written in the basis of $\{\ket{{\uparrow}(\theta_m)},\ket{{\downarrow}(\theta_m)}\}$ ($m=k,l$) as 
\begin{eqnarray}
H^{i,n}_{I_2}=i\frac{\eta_{n,i}\Omega_1\Omega_0}{4\Delta}  \left(\sigma_x^i \cos\theta_i +\sigma_z^i \sin\theta_i\right) b_n^\dagger e^{-i(\delta-\nu_n) t}   +h.c.
\end{eqnarray}
Thus, in the interaction picture with respect to Eq.\ref{second interaction} we obtain the effective Hamiltonian
\begin{eqnarray}
H_{{eff}_2}=\left(\frac{\Omega_0\Omega_1}{4\Delta}\right)^2\sum_{i,j,n} 
\frac{\eta_{n,i}\eta_{n,j}}{\delta-\nu_n}  \blb{\sigma_x^i  \sigma_x^j \cos\theta_i\cos\theta_j+ \frac{1}{2}\left( \sigma_z^i  \sigma_z^j+\sigma_y^i  \sigma_y^j \right) \sin\theta_i\sin\theta_j      },  \label{XXZ Hamiltonian}
\end{eqnarray}
if we assume that $\left(\frac{\Omega_0\Omega_1}{4\Delta}\right)^2\frac{\eta_{n,i}\eta_{n,j}}{\delta-\nu_n}  \ll \tilde{\Omega}_i = \tilde{\Omega}_l \ll \delta - \nu_n$. 
In order to suppress the leakage coming from an undesired $ \sigma_z^i  \sigma_z^j+\sigma_y^i  \sigma_y^j$ coupling, we can demand $\left(\frac{\Omega_0\Omega_1}{4\Delta}\right)^2\frac{\eta_{n,i}\eta_{n,j}}{\delta-\nu_n}  \ll\tilde{\Omega}_i - \tilde{\Omega}_l$, where $i,j$ are the notations of different blocks. The validity of the effective Hamiltonian is supported by our numerical simulation with realistic experiment parameters, as shown in Fig.\ref{fig-si:ion_sim}.

\begin{thebibliography}{99}

\providecommand{\url}[1]{\texttt{#1}}
\providecommand{\urlprefix}{URL }
\providecommand{\eprint}[2][]{\url{#2}}


\bibitem{Ladd10} T. D. Ladd, F. Jelezko, R. Laflamme, Y. Nakamura, C. Monroe, J. L. O'Brien, \href {http://www.nature.com/nature/journal/v464/n7285/full/nature08812.html}{{\it Nature} {\bf 464}, 45 (2010)}.
 
\bibitem{Cirac12} J. I. Cirac, P. Zoller, \href {http://www.nature.com/nphys/journal/v8/n4/full/nphys2275.html}{{\it Nature Physics} {\bf 8}, 264 (2012)}.

\bibitem{Nori14} I. M. Georgescu, S. Ashhab, Franco Nori, \href{http://journals.aps.org/rmp/abstract/10.1103/RevModPhys.86.153} {{\it Rev. Mod. Phys.} {\bf 86}, 153 (2014)}.

\bibitem{Jaksch14} T. H. Johnson, S. R. Clark, D. Jaksch, \href{http://www.epjquantumtechnology.com/content/1/1/10} {{\it EPJ Quantum Technology.} {\bf 1}, 10 (2014)}.

\bibitem{Grei02} M. Greiner, O. Mandel, T. Esslinger, T. W. H\"{a}nsch, and I. Bloch, \href{http://www.nature.com/nature/journal/v415/n6867/abs/415039a.html} {{\it Nature} {\bf 415}, 39 (2002)}.

\bibitem{Leib02} D. Leibfried, B. DeMarco, V. Meyer, M. Rowe, A. Ben-Kish, J. Britton, W. M. Itano, B. Jelenkovic, C. Langer,
T. Rosenband, and D. J. Wineland,  \href{http://journals.aps.org/prl/abstract/10.1103/PhysRevLett.89.247901} {{\it Phys. Rev. Lett.} {\bf 89}, 247901 (2002)}.

\bibitem{Frie08} A. Friedenauer, H. Schmitz, J. T. Gl\"{u}ckert, D. Porras, and T. Sch\"{a}tz, \href{http://www.nature.com/nphys/journal/v4/n10/abs/nphys1032.html} {{\it Nature Physics} {\bf  4}, 757 (2008)}.

\bibitem{Kim10} K. Kim, M.-S. Chang, S. Korenblit, R. Islam, E. E. Edwards, J. K. Freericks, G.-D. Lin, L.-M. Duan, and C. Monroe, \href{http://www.nature.com/nature/journal/v465/n7298/full/nature09071.html} {{\it Nature} {\bf 465}, 590 (2010)}.

\bibitem{Huelga97} S.F. Huelga, C. Macchiavello, T. Pellizzari, A.K. Ekert, M.B. Plenio and J.I. Cirac, \href{http://journals.aps.org/prl/abstract/10.1103/PhysRevLett.79.3865}{{\it Phys. Rev. Lett.} {\bf 79}, 3865 (1997)}.

\bibitem{Gio11} V. Giovannetti, S. Lloyd, L. Maccone, \href {http://www.nature.com/nphoton/journal/v5/n4/full/nphoton.2011.35.html}{{\it Nature Photonics} {\bf 5}, 222 (2011)}.


\bibitem{Kohler13} J. K\"{o}hler, J. A. J. M. Disselhorst, M. C. J. M. Donckers, E. J. J. Groenen, J. Schmidt, W. E. Moerner, \href{http://www.nature.com/nature/journal/v363/n6426/abs/363242a0.html}{{\it Nature} {\bf 363}, 242 (1993)}.

\bibitem{Wra13} J. Wrachtrup, C. von Borczyskowski, J. Bernard, M. Orritt, R. Brown, \href{http://www.nature.com/nature/journal/v363/n6426/abs/363244a0.html}{{\it Nature} {\bf 363}, 244 (1993)}.

\bibitem{Shor95} P. W. Shor, \href{http://journals.aps.org/pra/abstract/10.1103/PhysRevA.52.R2493}{{\it Phys. Rev. A} {\bf 52}, R2493 (1995)}.

\bibitem{Steane96} A. Steane, {\it Proc. Roy. Soc. Lond. A} {\bf 452}, 2551(1996).

\bibitem{Palma96} G. M. Palma, K.-A. Suominen and A. K. Ekert, {\it Proc. Roy. Soc. London Ser. A} {\bf 452}, 567 (1996).

\bibitem{Plenio97} M. B. Plenio, V. Vedral and P.L. Knight. \href{http://journals.aps.org/pra/abstract/10.1103/PhysRevA.55.67}{{\it Phys. Rev. A} {\bf 55}, 67 (1997)}.

\bibitem{Duan97} L.-M Duan and G.-C. Guo, \href{https://dx.doi.org/10.1103/PhysRevLett.79.1953}{{\it Phys. Rev. Lett.} {\bf 79} 1953 (1997)}.

\bibitem{Zanardi97} P. Zanardi and M. Rasetti, \href{http://journals.aps.org/prl/abstract/10.1103/PhysRevLett.79.3306}{{\it Phys. Rev. Lett.} {\bf 79}, 3306 (1997)}.

\bibitem{Lidar98} D. A. Lidar, I. L. Chuang, K. B. Whaley. \href{http://journals.aps.org/prl/abstract/10.1103/PhysRevLett.81.2594}{{\it Phys. Rev. Lett.} {\bf 81}, 2594 (1998)}.

\bibitem{Wu2002} L.-A. Wu, M. S. Byrd, and D. A. Lidar, \href{http://journals.aps.org/prl/abstract/10.1103/PhysRevLett.89.127901} {\it Phys. Rev. Lett.} {\bf 89}, 127901 (2002).

\bibitem{Lidar_review} D. A. Lidar, \href{http://onlinelibrary.wiley.com/doi/10.1002/9781118742631.ch11/summary} {\it Adv. Chem. Phys.} {\bf 154}, 295 (2014).

\bibitem{Hahn50} E. L. Hahn, \href{http://journals.aps.org/pr/abstract/10.1103/PhysRev.80.580} {{\it Phys. Rev.} {\bf 80}, 580 (1950)}.

\bibitem{Viola98} L. Viola, S. Lloyd, \href{http://journals.aps.org/pra/abstract/10.1103/PhysRevA.69.032314} {{\it Phys. Rev. A} {\bf 58}, 2733 (1998)}.

\bibitem{Pas04} P. Facchi, D. A. Lidar, S. Pascazio, \href{http://journals.aps.org/pra/abstract/10.1103/PhysRevA.69.032314} {{\it Phys. Rev. A} {\bf 69}, 032314 (2004)}.

\bibitem{Fan07} F. F. Fanchini, J. E. M. Hornos, R. d. J.~Napolitano, \href{http://journals.aps.org/pra/abstract/10.1103/PhysRevA.75.022329} {{\it Phys. Rev. A} {\bf 75}, 022329 (2007)}.

\bibitem{Lidar05} K. Khodjasteh, D. A. Lidar,  \href{http://journals.aps.org/prl/abstract/10.1103/PhysRevLett.95.180501} {{\it Phys. Rev. Lett.}  {\bf 95}, 180501 (2005)}.

\bibitem{Uhrig07} G. S. Uhrig,  \href{http://journals.aps.org/prl/abstract/10.1103/PhysRevLett.98.100504} {{\it Phys. Rev. Lett.} {\bf 98}, 100504 (2007)}.

\bibitem{Rabl09} P. Rabl, P. Cappellaro, M. V. Gurudev Dutt, L. Jiang, J. R. Maze, and M. D. Lukin, \href{http://journals.aps.org/prb/abstract/10.1103/PhysRevB.79.041302}{{\it Phys. Rev. B} {\bf 79}, 041302 (2009)}.

\bibitem{Gordon07} G. Gordon, N. Erez, G. Kurizki, J. Phys. B: At. Mol. Opt. Phys. 40 S75 (2007).

\bibitem{Tim11} N. Timoney, I. Baumgart, M. Johanning, A. F. Varon, M. B. Plenio, A. Retzker, Ch. Wunderlich, \href{http://www.nature.com/nature/journal/v476/n7359/full/nature10319.html}{{\it Nature} {\bf 476}, 185 (2011)}.

\bibitem{CaiNJP12} J.-M. Cai, B. Naydenov, R. Pfeiffer, L. P. McGuinness, K. D. Jahnke, F. Jelezko, M. B. Plenio, and A. Retzker, \href{http://dx.doi.org/10.1088/1367-2630/14/11/113023}{{\it New Journal of Physics} {\bf 14}, 113023 (2012)}.

\bibitem{Doh13} M. W. Doherty, N. B. Manson, P. Delaney, F. Jelezko, J. Wrachtrup, L. C.L. Hollenberg, \href{http://www.sciencedirect.com/science/article/pii/S0370157313000562}{\it Phys. Rep.} {\bf 528}, 1 (2013).

\bibitem{SI} See online supplemental material for the details of calculations.
 
\bibitem{Daniel96} Daniel T. Gillespie, \href{http://journals.aps.org/pre/abstract/10.1103/PhysRevE.54.2084}{{\it Phys. Rev. E} {\bf 54}, 2084 (1996)}.

\bibitem{Dolde13} F. Dolde, I. Jakobi, B. Naydenov, N. Zhao, S. Pezzagna, C. Trautmann, J. Meijer, P. Neumann, F. Jelezko, J. Wrachtrup, \href{http://www.nature.com/nphys/journal/v9/n3/full/nphys2545.html}{{\it Nature Physics} {\bf 9}, 139 (2013).}

\bibitem{Ozeri14nature} S. Kotler, N. Akerman, N Navon, Y. Glickman, R. Ozeri, \href{http://www.nature.com/nature/journal/v510/n7505/full/nature13403.html}{{\it Nature} {\bf 510}, 376 (2014).}

\bibitem{Albrecht14} A. Albrecht, G. Koplovitz, A. Retzker, F. Jelezko, S. Yochelis, D. Porath, Y. Nevo, O. Shoseyov, Y. Paltiel, M. B. Plenio, \href{http://iopscience.iop.org/1367-2630/16/9/093002/}{{\it New J. Phys.} {\bf 16}, 093002  (2014).}
  
 \bibitem{Britton12} J. W. Britton, Brian C. Sawyer, Adam C. Keith, C.-C. J. Wang, J. K. Freericks, H. Uys, M. J. Biercuk, J. J. Bollinger, \href{http://www.nature.com/nature/journal/v484/n7395/full/nature10981.html} {{\it Nature} {\bf 484}, 489 (2012)}.

\bibitem{Berprl11} A. Bermudez, J. Almeida, F. Schmidt-Kaler, A. Retzker and M. B. Plenio, \href{http://journals.aps.org/prl/abstract/10.1103/PhysRevLett.107.207209}{{\it Phys. Rev. Lett.} {\bf 107}, 207209 (2011)}.

\bibitem{Bernjp12} A. Bermudez, J. Almeida, K. Ott, H. Kaufmann, S. Ulm, F. Schmidt-Kaler, A. Retzker and M.B. Plenio, \href{http://iopscience.iop.org/1367-2630/14/9/093042/}{{\it New J. Phys.} {\bf 14}, 093042 (2012)}.

\bibitem{Islam13} R. Islam, C. Senko, W. C. Campbell, S. Korenblit, J. Smith, A. Lee, E. E. Edwards, C. C. J. Wang, J. K. Freericks, C. Monroe, \href{http://www.sciencemag.org/content/340/6132/583.abstract}{{\it Science} {\bf 340}, 583 (2013)}.

\bibitem{Monz09} T. Monz, K. Kim, A. S. Villar, P. Schindler, M. Chwalla, M. Riebe, C. F. Roos, H. H\"{a}ffner, W. H\"{a}nsel, M. Hennrich, R. Blatt, \href{http://journals.aps.org/prl/abstract/10.1103/PhysRevLett.103.200503} {\it Phys. Rev. Lett.} {\bf 103}, 200503 (2009).

\bibitem{Cardy07} P. Calabrese, J. Cardy, \href{http://iopscience.iop.org/1742-5468/2007/06/P06008/} {{\it J. Stat. Mech.} {\bf 0706}, P06008 (2007)}.

\bibitem{Rigol08} M. Rigol, V. Dunjko, M. Olshanii, \href{http://www.nature.com/nature/journal/v452/n7189/abs/nature06838.html} {{\it Nature} {\bf 452}, 854 (2008)}. 

\bibitem{Riera12} A. Riera, C. Gogolin, J. Eisert,  \href{http://journals.aps.org/prl/abstract/10.1103/PhysRevLett.108.080402} {{\it Phys. Rev. Lett.} {\bf 108}, 080402 (2012)}.

\bibitem{Trotzky12} S. Trotzky, Y.-A. Chen, A. Flesch, I. P. McCulloch, U. Schollw\"{o}ck, J. Eisert, I. Bloch, \href{http://www.nature.com/nphys/journal/v8/n4/full/nphys2232.html} {{\it Nature Physics} {\bf 8}, 325 (2012)}.

\bibitem{Porras04} D. Porras and J. I. Cirac, \href{http://journals.aps.org/prl/abstract/10.1103/PhysRevLett.92.207901} {{\it Phys. Rev. Lett.}  {\bf 92}, 207901 (2004)}.

\bibitem{Zhu06} S.-L. Zhu, C. Monroe, and L.-M. Duan, \href{http://journals.aps.org/prl/abstract/10.1103/PhysRevLett.97.050505} {{\it Phys. Rev. Lett.}  {\bf 97}, 050505 (2006)}.

\bibitem{Ber12} A. Bermudez, P. O. Schmidt, M. B. Plenio, and A. Retzker, \href{http://journals.aps.org/pra/abstract/10.1103/PhysRevA.85.040302}{\it Phys. Rev. A} {\bf 85}, 040302(R) (2012).

\bibitem{Itsik1} I. Cohen and A. Retzker \href{http://journals.aps.org/prl/abstract/10.1103/PhysRevLett.112.040503}{Phys. Rev. Lett. {\bf 112,} 040503 (2014).}

\bibitem{Itsik2} I. Cohen, P. Richerme, Z.-X. Gong, C. Monroe, A. Retzker, \href{http://arxiv.org/abs/1505.04695} {arXiv:1505.04695.}

\end{thebibliography}

\begin{thebibliography}{99}

\providecommand{\url}[1]{\texttt{#1}}
\providecommand{\urlprefix}{URL }
\providecommand{\eprint}[2][]{\url{#2}}

 
\bibitem{Daniel96_SI} Daniel T. Gillespie, \href{http://journals.aps.org/pre/abstract/10.1103/PhysRevE.54.2084}{{\it Phys. Rev. E} {\bf 54}, 2084 (1996)}.

\bibitem{Dolde13_SI} F. Dolde, I. Jakobi, B. Naydenov, N. Zhao, S. Pezzagna, C. Trautmann, J. Meijer, P. Neumann, F. Jelezko, J. Wrachtrup, \href{http://www.nature.com/nphys/journal/v9/n3/full/nphys2545.html}{{\it Nature Physics} {\bf 9}, 139 (2013).}

\bibitem{Ber12_SI} A. Bermudez, P. O. Schmidt, M. B. Plenio, and A. Retzker, \href{http://journals.aps.org/pra/abstract/10.1103/PhysRevA.85.040302}{{\it Phys. Rev. A} {\bf 85}, 040302(R) (2012)}.

\bibitem{Itsik1} I. Cohen and A. Retzker \href{http://journals.aps.org/prl/abstract/10.1103/PhysRevLett.112.040503}{Phys. Rev. Lett. {\bf 112,} 040503 (2014).}

\bibitem{Itsik2} I. Cohen, P. Richerme, C. Monroe, and A. Retzker, to be published.

\end{thebibliography}
\end{document}